\documentclass[twocolumn]{aastex63}

\shorttitle{Active Region Emergence}
\shortauthors{Rees-Crockford et al.}

\graphicspath{{./}{figures/}}

\begin{document}

\title{Pre-emergence Signatures Of Horizontal Divergent Flows In Solar Active Regions}

\correspondingauthor{T. Rees-Crockford}
\email{t.reescrockford@qub.ac.uk}

\author[0000-0003-4243-1776]{T. Rees-Crockford}
\affiliation{Astrophysics Research Centre, School of Mathematics and Physics, Queen's University Belfast, Belfast BT7 1NN, UK}

\author[0000-0003-1400-8356]{C. J. Nelson}
\affiliation{Astrophysics Research Centre, School of Mathematics and Physics, Queen's University Belfast, Belfast BT7 1NN, UK}
\affiliation{European Space Agency (ESA), European Space Research and Technology Centre (ESTEC), Keplerlaan 1, 2201 AZ, Noordwijk, The Netherlands}

\author[0000-0002-7725-6296]{M. Mathioudakis}
\affiliation{Astrophysics Research Centre, School of Mathematics and Physics, Queen's University Belfast, Belfast BT7 1NN, UK}

\begin{abstract}
Solar active regions (ARs) play a fundamental role in driving many of the geo-effective eruptions which propagate into the Solar System. However, we are still unable to consistently predict where and when ARs will occur across the solar disk by identifying pre-emergence signatures in observables such as the Doppler velocity (without using Helioseismic methods). Here we aim to determine the earliest time at which pre-emergence signatures, specifically the Horizontal Divergent Flow (HDF), can be confidently detected using data from the Solar Dynamics Observatory's Helioseismic and Magnetic Imager (SDO/HMI). Initially, we follow previous studies using the thresholding method, which searches for significant increases in the number of pixels that display a specific line-of-sight velocity. We expand this method to more velocity windows and conduct a basic parameter study investigating the effect of cadence on the inferred results. Our findings agree with previous studies with $37.5$\% of ARs displaying a HDF, with average lead times between the HDF and flux emergence of $58$ minutes. We present a new potential signature of flux emergence which manifests as cadence-independent transient disruptions to the amplitudes of multiple velocity windows and recover potential pre-emergence signatures for 10 of the 16 ARs studied, with lead times of 60-156 minutes. Several effects can influence both the estimated times of both HDF and flux emergence suggesting that one may need to combine Doppler and magnetic field data to get a reliable indicator of continued flux emergence.
\end{abstract}


\section{Introduction} \label{sec:intro}

Active regions (ARs) are locations where large amounts of magnetic flux break through the solar surface from the solar interior and is thought to occur due to buoyancy acting upon flux \citep{1955ApJ...121..491P, 1979SoPh...62...23A} that originates within the deep convection zone. Once sufficient flux breaks through the solar surface, it typically forms into a bipole and a surrounding plage, with the regions of negative and positive fields aligning with Hale's and Joy's laws \citep{1919ApJ....49..153H} (in reality a huge variety of different permutations exist in terms of AR structure). In addition to the detection of flux in magnetograms, velocity fields accompanying the newly emerging magnetic field have also been identified \citep[and references therein]{1976SoPh...46..125K}, corresponding to the bulk movement of plasma both contained within and around the magnetic flux. Helioseismological techniques have been used to determine that velocity fields could also be recovered pre-emergence (\citealt{1995ASPC...76..250B,2008SoPh..251..369Z,2011Sci...333..993I,2013ApJ...762..131B}) indicating that it may be possible to predict AR emergence, though the recovered velocities and time lead varied with AR and technique. More recently, \citet{2012ApJ...751..154T, 2014ApJ...794...19T} identified signatures of pre-emergence (referred to as Horizontal Divergent Flows, or HDFs) in simple Doppler measurements sampled by the Solar Dynamics Observatory's Helioseismic and Magnetic Imager (SDO/HMI; \citealt{Scherrer12}), however, their statistical results were limited to HDFs detected in a single velocity window between [$1$ km s$^{-1}$, $1.5$ km s$^{-1}$]. In this work, we extend the results of \citet{2012ApJ...751..154T} to a larger range of velocity windows and discuss the effects of cadence on the methods used.

Much of our understanding of AR emergence has been developed through analysis of realistic numerical simulations over the past few decades. It is now thought that emergence could take place in two steps, with the first being the flux tube rising to just below the solar surface, and the second being when the flux tube actually emerges into the atmosphere (\citealt{2004A&A...426.1047A}). As part of this second step, the flux tube would expand laterally as the plasma above it would impede its upward motion. Flows associated with this lateral expansion have been seen in simulations (\citet{2007A&A...467..703C, 2010ApJ...720..233C}), where the lateral expansion of the flux scaled with the magnetic field and the plasma density such that $B\sim\rho^{1/2}$. It was noted by \citet{2010ApJ...714..505T} that deceleration of the emerging flux in the chromosphere, rather than the upper convection zone (i.e. beneath the surface), may occur due to the structure of the emerging flux, thereby allowing faster draining of the plasma and thus a faster emergence. These authors considered the emergence of said flux sheet from a depth of 20~Mm below the surface, and suggested a threshold for two-step emergence (over failed emergence) at $10^{21}$--$10^{22}$~Mx.

Further 3D work by \citet{2013A&A...553A..55T} studied the effect that varying the field strength, twist, and the radius of curvature of a flux tube has on flux emergence. They found that the rise speed of the flux was strongly dependent on the magnetic field strength, but only weakly on the twist and radius of the tube. In their strong field cases, they found that the increased field strength led to faster emergence, and thus faster but shorter HDFs as the plasma has less time to drain before emergence. No correlation was found between the duration of the HDF and the twist, but they did note that higher twists did also have a positive correlation with the maximum speed of the HDF. They suggested that this was due to the twist allowing the flux tube to remain more intact during the emergence, thereby allowing more of it to emerge at once. This work was expanded upon by \citet{2019ApJ...874...15S}, who also investigated the length of the emerging portion of the flux tube, and the scaling of $B\sim\rho^{\kappa}$. They found that there were also geometric parameters that could define whether flux would successfully emerge, with a larger tube radius causing more rapid emergence. Furthermore, tubes with shorter buoyant sections emerged more slowly than those with longer buoyant sections, with emergence not being possible below a certain length (scaled to other parameters, see their Sec. 3.3.3).

Observational evidence of the flows predicted by these numerical simulations has also been acquired over recent years using simple Dopplergrams. 
For instance, \citet{2007AstL...33..766G} used SOHO/MDI data to determine the line-of-sight (LOS) velocity field of an emerging AR. Using these observations, they found upwards velocities with a mean of -0.23~km~s$^{-1}$, and a maximum of $\sim$2~km~s$^{-1}$ over the first 2 hours of the emergence. Although they only studied data preceding the emergence by 3 hours, they also noted the appearance of velocities around -0.4~km~s$^{-1}$ just before the emergence of the first magnetic flux. They attributed these motions to be the emergence of the flux tube through the photosphere. Early observational evidence of two-step flux emergence have been reported by  \citet{2011PASJ...63.1047O}, who found the apparent deceleration and spreading of flux as it emerged into the chromosphere. The first observational evidence of the lateral flows reported by \citet{2010ApJ...720..233C} was presented by \citet{2012ApJ...751..154T}. Those authors studied an AR close to the solar limb, reasoning that at a sufficient angle, one should be able to see the horizontal component of these motions in the wings of LOS Doppler profiles before flux emergence. Following their analysis, \citet{2012ApJ...751..154T} found significant increases in the number of pixels which displayed horizontal velocities around 0.6-1.5~km~s$^{-1}$ preceding the emergence by 100 minutes. Additional evidence of these flows was found in the follow up statistical work of \citet{2014ApJ...794...19T}, who found HDFs in 13 of 23 events. They found an average maximum HDF velocity (see their Sec.~3.2) of $\sim$3.1~km~s$^{-1}$, and an average time difference of 61 minutes between the on-set of these motions and the emergence of flux. Observationally, we must of course be careful to distinguish what different authors mean by emergence. In some articles, emergence is used to refer to the initial breach of flux into the photosphere, whereas other authors use it to describe the entire time-period before the AR reaches its maximum flux. Here we define emergence as the initial breach of the flux above a threshold value in order to provide comparable results with \citet{2012ApJ...751..154T,2014ApJ...794...19T}.

\begin{figure*}
    \centering
    \includegraphics[width=\textwidth]{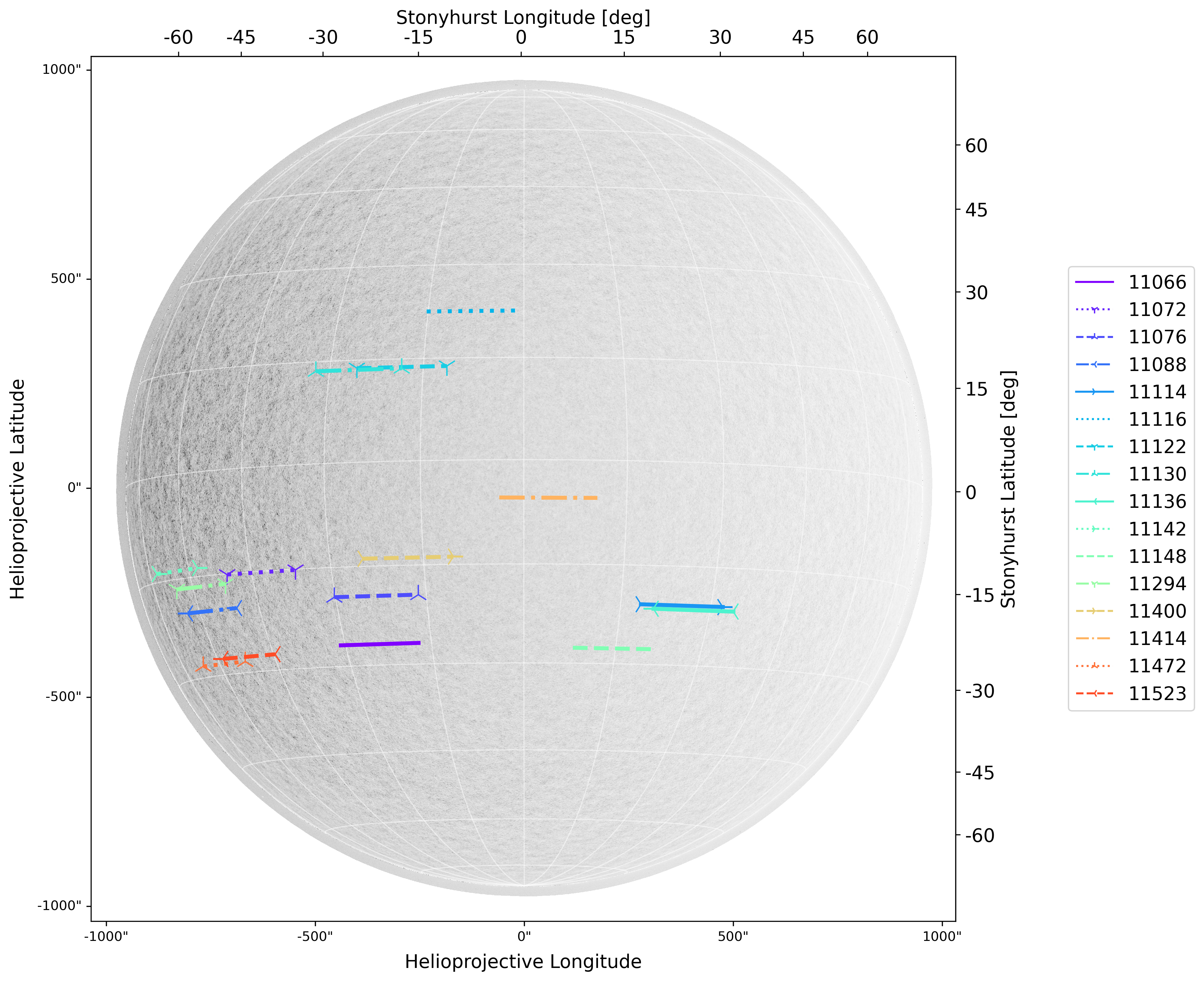}
    \caption{Location of each AR analysed here during the studied time-period. The colour, line style, and marker corresponding to each AR are included in the legend on the right-hand side.}
    \label{fig:ar_track_locs}
\end{figure*}

In this article, we build on the work of \citet{2012ApJ...751..154T,2014ApJ...794...19T} in order to investigate whether pre-emergence HDF signatures can be detected in a wider variety of velocity bins. Additionally, we study the effect of cadence on our ability to detect a clear HDF, thereby investigating whether this method is scalable between different instruments. Our work is structured as follows: In Section~\ref{sec:methods} we detail our AR selection and data processing methods; in Section~\ref{sec:results} we present our results; in Section~\ref{sec:disc} we discuss our results in the context of the literature; we draw our conclusions in Section~\ref{sec:concs}.

\section{Methods} \label{sec:methods}

\subsection{Data Selection and Processing}\label{sec:data_selec}

\begin{deluxetable}{ccccccc}[ht]
\tablecaption{Properties Of The 16 P0 Active Regions Studied. \label{tab:ar_list}}
\tablewidth{0pt}
\tablehead{
\colhead{NOAA}& \colhead{HARP} & \colhead{Schunker} &\colhead{Lon.} 
&\colhead{Lon.} &\colhead{Lat.}\\
\colhead{\#} & \colhead{T\_FRST1  \tablenotemark{a}} & \colhead{$T_{S}$} &\colhead{$T_{S}$\tablenotemark{b}}&\colhead{$T_{E}$\tablenotemark{c}}&\colhead{$T_{S}$\tablenotemark{d}}}
\startdata
11066 & 2010.05.02\_23:12 & 23:48 &-30.9 &-17.4 &-26.7 \\
11072 & 2010.05.20\_16:24 & 17:12 & -50.4 &-36.7 &-15.1 \\
11076 & 2010.05.31\_04:12 & 06:24 & -30.2 &-17.4 &-19.4 \\
11088 & 2010.07.11\_07:48 & 08:36 & -64.0 &-50.4 &-20.1 \\
11114 & 2010.10.14\_04:12 & 04:12 &  18.3 & 32.3 &-20.8 \\
11116 & 2010.10.16\_19:24 & 22:48 & -15.2 & -03.2 & 22.3 \\
11122 & 2010.11.06\_00:48 & 01:12 & -25.5 &-11.3 & 13.8 \\
11130 & 2010.11.27\_15:12 & 18:12 & -32.4 &-19.9 & 13.6 \\
11136 & 2010.12.24\_07:48 & 08:24 &  20.4 & 34.0 &-21.4 \\
11142 & 2010.12.31\_09:00 & 09:24 & -71.5 &-57.4 &-13.8 \\
11148 & 2011.01.17\_02:00 & 02:24 &  08.0 & 21.5 &-27.6 \\
11294 & 2011.09.11\_01:24 & 04:12 & -65.3 &-52.8 &-16.4 \\
11400 & 2012.01.14\_02:00 & 02:00 & -24.5 &-10.3 &-13.9 \\
11414 & 2012.02.04\_09:24 & 09:24 & -03.5 & 10.8 &-05.4 \\
11472 & 2012.04.29\_04:12 & 05:24 & -66.1 &-53.1 &-28.3 \\
11523 & 2012.07.11\_23:12 & 23:24 & -58.4 &-44.5 &-27.5 \\
\enddata

\tablenotetext{a}{Time of initial detection in HARP.}
\tablenotetext{b}{HARP LON\_FWT differentially rotated to start of data set.}
\tablenotetext{c}{LON\_FWT at Schunker start time.}
\tablenotetext{d}{As b, for latitude.}

\end{deluxetable} 

In order to study the LOS magnetic field strengths and Doppler velocities of ARs in the solar photosphere, we use data obtained by the SDO/HMI instrument (using the 6173~\AA\ line).
The data is provided in the form of full-disk magnetograms (hmi.M\_45s) and Dopplergrams (hmi.V\_45s) with a spatial resolution of 1\arcsec\ and a $45$ s cadence.
These parameters are sufficient for studying large-scale, long-lived solar phenomena such as ARs. 
We use data keywords provided by the  Space-weather HMI Active Region Patches (SHARP) hmi.sharp\_720s series \citep{2007AN....328..352B} in order to identify the spatial locations of ARs, used for derotating the data to the relevant times. 
For our subsequent analysis, we choose the ARs defined as ``P0'' (``emergence into a very quiet region'') by \citet{2016A&A...595A.107S} as our targets. 
Of the 21 ARs defined as P0 in that work, 16 had data available at the time of our analysis. 
In Table \ref{tab:ar_list}, we provide the basic information for these ARs, including timings and relevant spatial co-ordinates. 
Note that all times presented here are in TAI \citep{si-brochure}. 
Although \citeauthor{2016A&A...595A.107S} defined their emergence time as ``the time when the absolute flux, corrected for line-of-sight projection, reaches 10\% of its maximum value over a 36-h interval following the first appearance of the sunspot (or group) in the NOAA record'', we, as previously stated, define the emergence time as the first point at which the number of pixels within a specific magnetic field strength window increases over a threshold value of $1\sigma$ in order to better align our work with that of \citet{2012ApJ...751..154T}. 
In Fig.~\ref{fig:ar_track_locs} we show the tracked location of each AR studied here.

\begin{figure}[t]
    \centering
    \includegraphics[width=0.45\textwidth,trim =6cm 3cm 5cm 1.5cm]{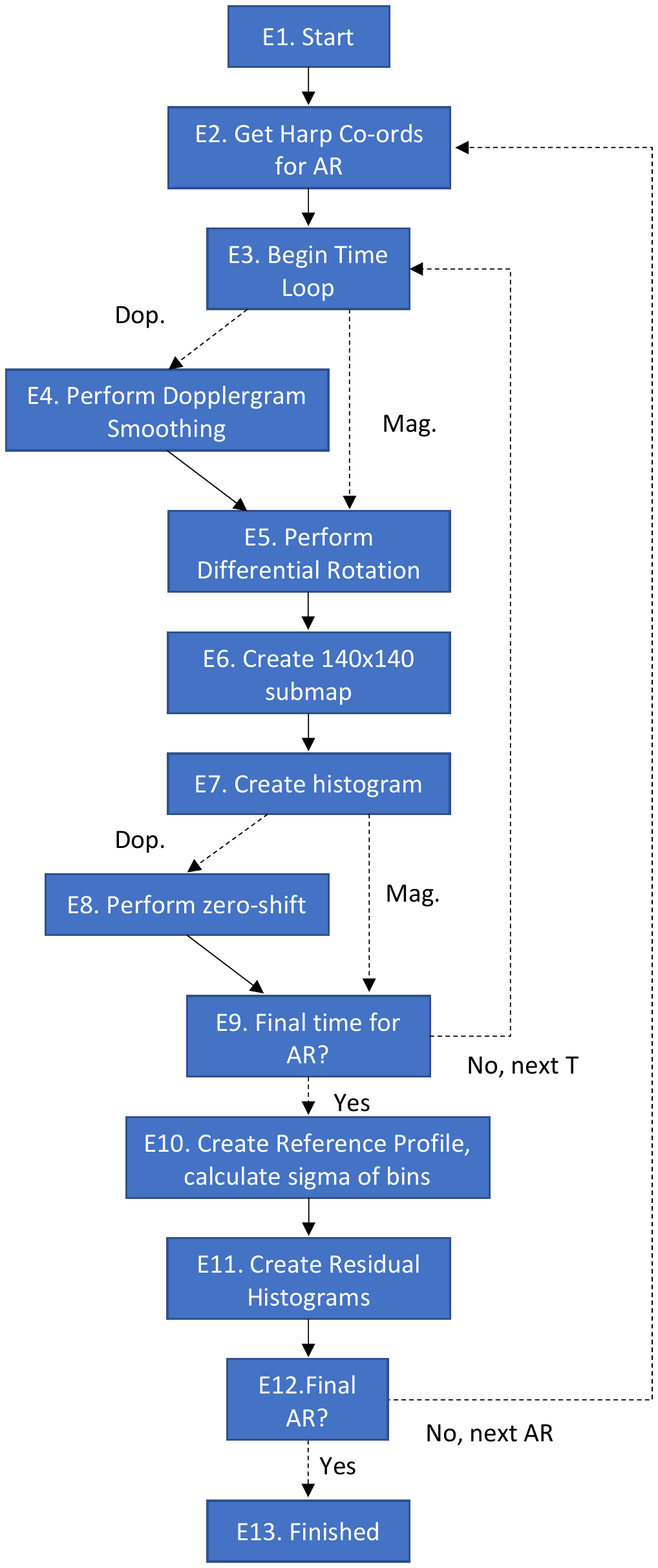}
    \caption{Flowchart of the data extraction method for both the magnetic field strength and Doppler velocity datasets.}
    \label{fig:flowchart}
\end{figure}

In this article, we track the number of pixels with specific magnetic field strengths and Doppler velocities within a given Region of Interest (ROI).
The ROI is defined by a $140$\arcsec$\times140$\arcsec\ box centred on the HARP keywords of LON\_FWT and LAT\_FWT (in Stonyhurst co-ordinates), which are are defined as the flux weighted centre of the AR at the time of the initial detection as recorded in the HARP keyword T\_FRST1 (defined as the first recorded time of this HARP number).
These coordinates are then differentially rotated back to the appropriate position at each sampled time. 
The data used here cover the 24-hours period preceding that of the Schunker emergence time for each event.
We use either a $45$ second or a $12$ minute cadence depending on the section, with the studied cadence being noted explicitly for each result.
We use the $45$~s cadence on the 3 ARs common to both this analysis and \citet{2014ApJ...794...19T} to allow for comparison between them.
This also allows us to identify if any differences between the methods could cause significant differences in the results.
This is detailed further in Sec.~\ref{sec:tor_com}.
The lower $12$ minute cadence is chosen to both conduct a basic parameter study about the effect of the cadence on the method and due to reasons of data storage and processing speed. 
Our data extraction method, based on that of \citet{2012ApJ...751..154T} with some differences, is performed using Python (Numpy: \citet{harris2020array}, Sunpy: \citet{sunpy_community2020}, Astropy: \citet{astropy:2013}, \citet{astropy:2018}).
The basic outline of the method is included in Fig.~\ref{fig:flowchart}, with this figure being referenced throughout the following method: 
\begin{enumerate}
    \item We begin by retrieving the HARP co-ordinate keywords for the relevant AR (E2). 
    \item We then perform a running average from the Dopplergram data sampled over $48$ minutes (E4). Unlike \citet{2012ApJ...751..154T} who do a running average over 30 minutes, we average over 48 minutes in order to maintain consistency between the higher ($45$ s) and lower ($12$ minute) cadences considered in our analysis.
    \item The co-ordinates for each time-step are then differentially rotated to the appropriate time (E5), and the $140$\arcsec$\times140$\arcsec\ submap of the ROI is created (E6). Unlike in \citet{2012ApJ...751..154T,2014ApJ...794...19T}, here we do not perform a Postel projection. This is detailed further below.
    \item A histogram is taken of the data for each time-step (E7). In the case of the Dopplergram, this histogram is shifted to have its peak at zero (E8). This is done by subtracting the radial velocity of the satellite ( as defined in the FITS header keyword OBS\_VR), and then the mean velocity of the histogram to mitigate part of the effect of the satellite motion. This deviates from the method used by \citet{2014ApJ...794...19T}, and will be discussed below. For magnetograms, we take the histogram of the absolute value of the signed flux. 
    \item A reference profile histogram is created using the average of the first 3 hours worth of data (E10). The standard deviation, $\sigma$, in each histogram bin across that time is then calculated (E10). 
    \item The reference profile is subtracted from each histogram across the entire time range to create residual profiles, which are the profiles presented in this work (E11). 
\end{enumerate}

As mentioned in Step 3, we do not perform a Postel projection.
This is primarily because we wish to analyse the data whilst minimising any potential artifacts introduced by processing. 
Additionally, as mentioned in Step 4, we deviate slightly from the reduction method used by \citet{2014ApJ...794...19T}, as outlined in their Sec. 2.2, as again we wish to minimise the number of post-processing steps applied to the data. 
In that sense, we believe that it is more appropriate to simply subtract the mean as is performed in \citet{2012ApJ...751..154T}. 
There is also another issue in Step 4 that we must address.
Whilst we attempt to account for the orbital motion of the satellite, we cannot fully remove it.
This is primarily due to underlying systemic reasons \citep[see Sec. 3.1 in][]{2016ApJ...823..101S}, and the lack of tested and verified libraries within sunpy.
Furthermore, the Sun's rotation creates an asymmetry between the East and West limbs, as one moves towards and away from the observer, proportional to the differential rotation at any given latitude. 
As this trend reverses across the disk, we must also be aware of the effect when we take the reference profile. 
Furthermore, the non-radial components of the motion of the satellite and additional instrumental effects \citep{2016ApJ...823..101S} will further modify the underlying distribution and thus, how the subsequent residual profiles will evolve. 
For this work, these effects manifest as follows: firstly, there is an underlying oscillatory behaviour to the Doppler profiles; and secondly, as the longitude of the studied AR changes relative to its reference profile, it will experience inherent changes to the residual profile.
Ultimately, however, the underlying trend will provide the backdrop against which we are looking, but will not dictate whether we are capable of seeing the transient features we are looking for.
This is because the satellite motion will vary smoothly over the course of the observation (24 hours), and will have only minor effects over the time ranges considered for the transient changes studied here (0-3 hours).
In addition, we believe that the removal of the radial component of the satellite motion and the zero-shift (see point 4 in the method outline) of the histogram results in velocity profiles are sufficient for this analysis.
In addition, these profiles have been smoothed using the \texttt{numpy.convolve} function of width 5 in ``valid'' mode (M4, D4).
This smoothing has been applied to avoid biasing the method against cases where the profile dips below the threshold for only one frame. 
Note that this smoothing is applied to the resultant profiles, not the data itself.

\begin{figure*}[!t]
    \centering
    \includegraphics[width=\textwidth,trim = 0cm 8cm 0cm 2cm]{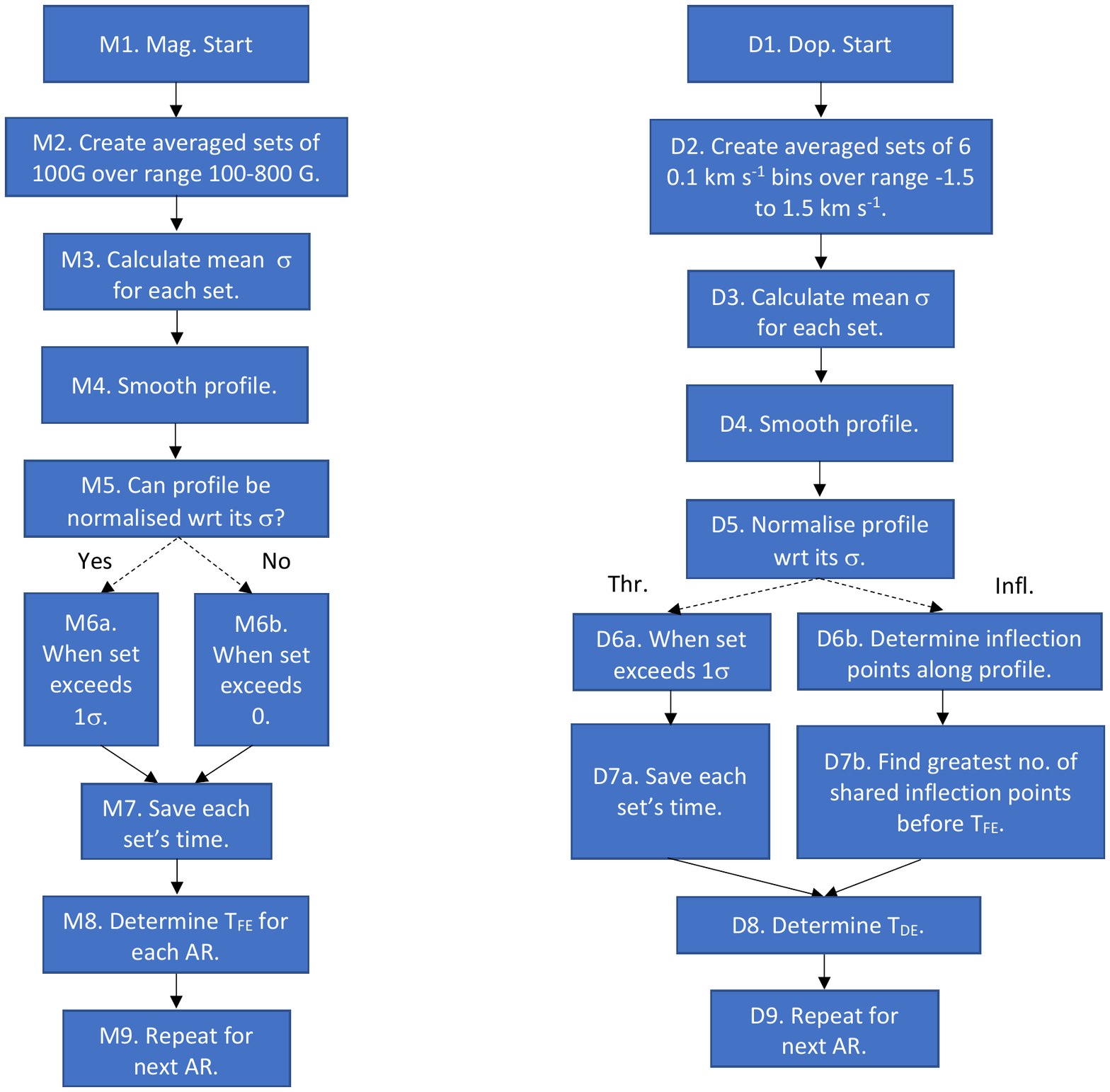}
    \caption{(Left panel) The flow chart of the determination of the time of magnetic emergence, $T_{FE}$. (Right panel) The flow chart of the determination of the Doppler time of pre-emergence, $T_{DE}$ for both methods used here. Note that for all cases, the profiles used here are those of the residual profiles discussed in Fig.~\ref{fig:flowchart}.}
    \label{fig:mag_dop_flowcharts}
\end{figure*}

\subsection{Creating Data Products For Analysis}\label{sec:data_an}

The purpose of our analysis is to determine the earliest time at which we can detect HDF signatures in Dopplergrams.
In order to do this we must first determine the time of magnetic flux emergence, $T_{FE}$ (see left side of Fig.~\ref{fig:mag_dop_flowcharts}).
As previously discussed in \citet{2012ApJ...751..154T}, the time of emergence is defined as the time at which the number of pixels which display magnetic field strengths within a specific window exceeds the averaged standard deviation of that set of bins calculated from the time-period used to create the reference profile.
For our magnetic profiles we create bins with widths of 100~G (i.e over the interval [lower bound, lower bound+100) G), over the range 100 to 800~G  (using the absolute value of the magnetic field). 
We exclude the [0, 100) G set due to the inherent noise hence the large variance and standard deviation.
Once $T_{FE}$ has been determined from the relevant magnetic profiles, we then begin searching the Doppler profiles for an HDF, which has a time denoted by $T_{DE}$, using one of two methods (see right side of Fig.~\ref{fig:mag_dop_flowcharts}).

In order to study the evolution of the Doppler velocity within the ROI, we follow a method similar to that applied to the magnetogram data by first summing the number of pixels within velocity bins which cover the interval [lower bound, lower bound+0.1) km~s$^{-1}$. 
Velocity windows are then created, as in \citet{2014ApJ...794...19T}, mirrored around $0$ km s$^{-1}$, each with $6$ bins. 
This differs slightly from the work of \citet{2014ApJ...794...19T} who used 8 bins per window.
The specific velocity windows studied at any time are indicated in all relevant figures (see, for example, the legend of Fig.~\ref{fig:com_full_res}). 
We chose $6$ bins instead of $8$ as this increases the number of windows we can compare against each other and gives us more information at lower Doppler velocities.
 In this work we do not explicitly distinguish between the upwards velocity of the emerging flux, V$_z$, and the horizontal velocity of the divergent flow, V$_h$ to use the nomenclature of \citet{2012ApJ...751..154T} (see their Fig.~8). 
This is for two reasons. 
Firstly, and related to the lack of a Postel projection, is because we wish to minimise the introduction of any processing artifacts.
Secondly, whilst we do not explicitly distinguish between the vertical and horizontal velocity components, we nonetheless deal with them implicitly. 
This is done through the consideration of all velocity windows, and the assumption that either there will be a recoverable component of the emerging flux at any longitude, or that the emergence will cause sufficient disruption to the underlying distribution to be measurable.

It should be noted that the flowcharts presented in Fig.~\ref{fig:mag_dop_flowcharts} assume a ``successful'' determination of $T_{DE}$ in at least one Doppler velocity window.
In the case that this is not possible, no times are recorded for that AR. 
This is not of concern here as there are a sufficient number of windows in both the magnetograms and Dopplergrams to allow the determination of $T_{DE}$ for each AR, and as such it will not be considered further.
In order to better outline our method for the reader, and to provide a more direct comparison to the work of \citet{2012ApJ...751..154T,2014ApJ...794...19T}, we include an example of this method applied to full cadence ($45$ s) data in Sec.~\ref{sec:tor_com}.

\section{Results}\label{sec:results}

\subsection{Method Comparison}\label{sec:tor_com}

\begin{figure*}
    \centering
    \includegraphics[width=\textwidth]{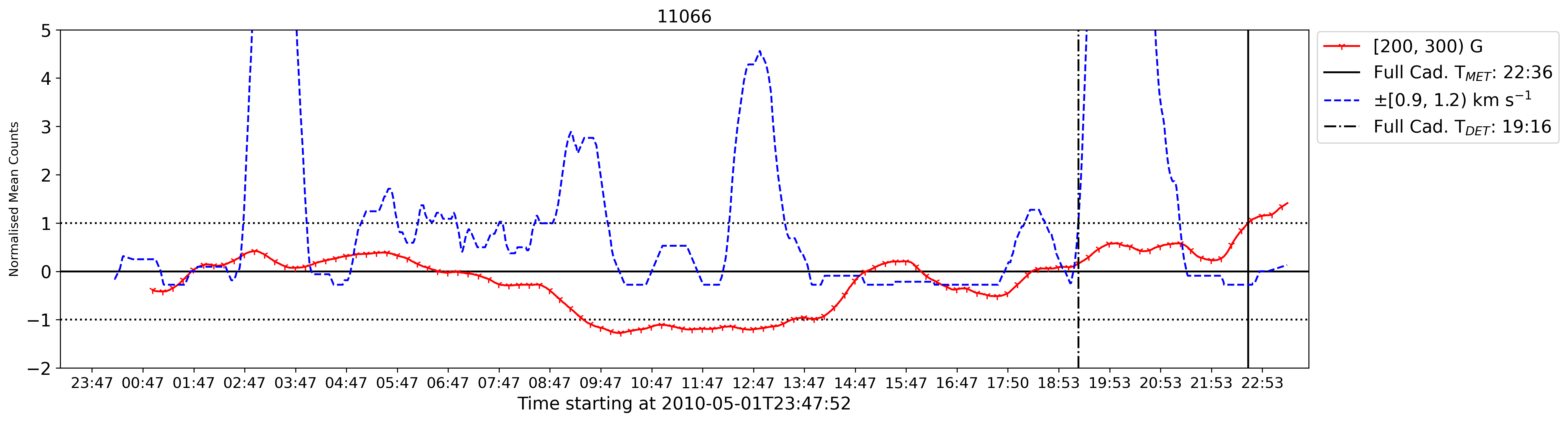}
    \includegraphics[width=\textwidth]{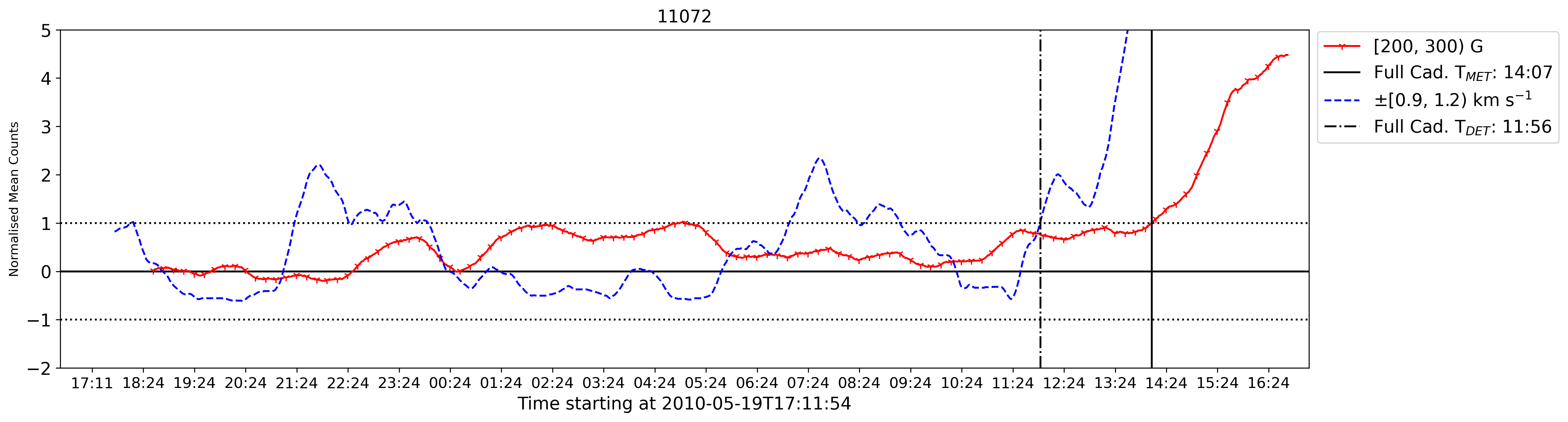}
    \includegraphics[width=\textwidth]{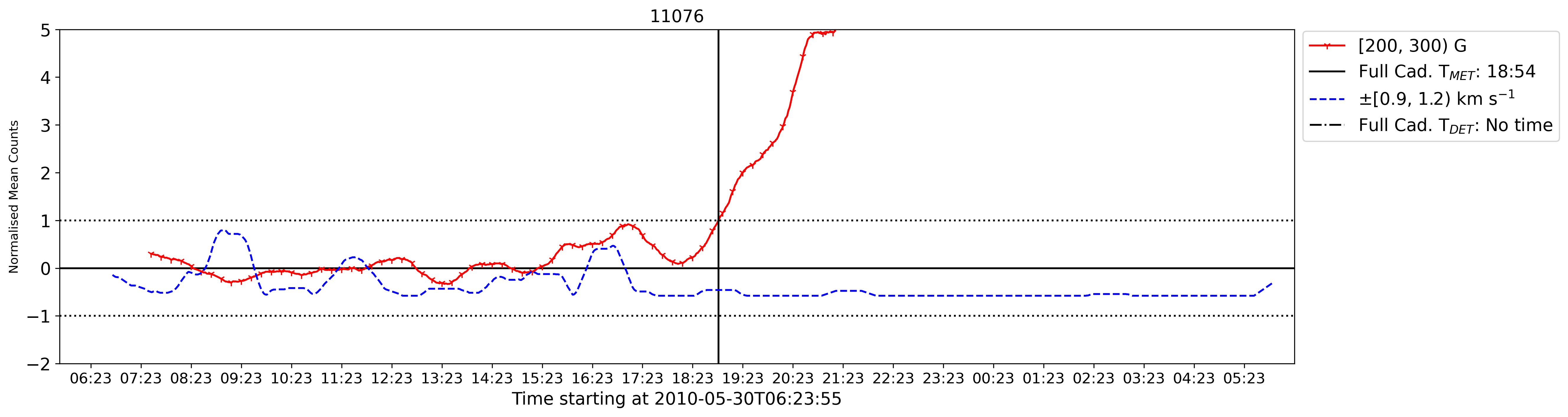}
    \caption{Evolution of the relevant magnetic field (red lines) and Doppler velocity (dashed blue lines) windows for the three ARs compared to \citet{2014ApJ...794...19T} (as discussed in Sect.~\ref{sec:tor_com}). The counts have been normalised against the standard deviations calculated from the initial 3-hour background time. The flux emergence above $1\sigma$ is clear in each event (denoted by the vertical solid black lines), with the presence of the HDF also being apparent in the Doppler data for ARs 11066 and 11072 (denoted by the vertical dot-dashed black lines). No HDF is detected for AR 11076.}
    \label{fig:com_full_res}
\end{figure*}

We initially apply our method to three ARs using full-cadence ($45$ s) magnetic field and Dopplergram data.
The three ARs chosen (ARs 11066, 11072, and 11076) were also studied by \citet{2014ApJ...794...19T} meaning we are able to identify whether the slight differences in our methodology introduce any major changes to the results. 
For these three ARs, \citet{2014ApJ...794...19T} found $T_{FE}$ values of $20$:$30$ UT, $14$:$51$ UT, and $16$:$31$ UT, respectively, for the magnetic field strength window spanning [$200$, $300$) G (assuming these are `typical' ARs).
Through our analysis, we find values of $22$:$35$ UT, $14$:$06$ UT, and $18$:$53$ UT, respectively, from the same magnetic field strength window. 
We do expect to report slightly different emergence times due to the minor variations in our methodology, and we consider these differences of around $2$ hours to be acceptable (given the time-scales of the processes studied here).
For $T_{DET}$, \citet{2014ApJ...794...19T} reported values of $19$:$25$ UT, $13$:$05$ UT and $17$:$08$ UT (note this is after $T_{FE}$ and, as such, no HDF was reported) for a velocity window spanning $\pm$[1, 1.5) km s$^{-1}$.
From our analysis, and using a velocity window spanning $\pm$[$0.9$, $1.2$) km s$^{-1}$ we find values of $19$:$16$ UT and $11$:$56$ UT for ARs 11066 and 11072, but find no HDF for AR 11076, agreeing with the results of \citet{2014ApJ...794...19T}.
Overall, we are confident that our method is providing accurate results  that are comparable to the previous literature despite the slight changes to the methodology, which will allow us to study the presence of HDF signatures in a wider range of velocity windows. 

We plot the evolution of the reported magnetic and velocity windows at full-cadence in Fig.~\ref{fig:com_full_res} for each of the three ARs.
The red lines denote the standard deviation normalised number of pixels within the appropriate magnetic field strength window, whilst the dashed blue lines plot the same for the velocity window.
The horizontal dashed lines denote the $1\sigma$ level used to define the threshold values in this study. 
For AR 11066 (top panel), the HDF is clearly present (indicated by the vertical dot-dashed black line) as the large rise in the dashed blue line several hours before the emergence of the flux (which takes place right at the end of the studied time-period; denoted by the solid vertical black line). 
For AR 11072 (middle panel), the HDF begins only a few minutes before the emergence of the flux with the dashed blue and red lines seemingly increasing around the same time. 
For AR 11076 (bottom panel), the flux clearly begins to emerge around half way through the studied time-period, but no associated HDF is detected.

\begin{figure*}
    \centering
    \includegraphics[width=\textwidth]{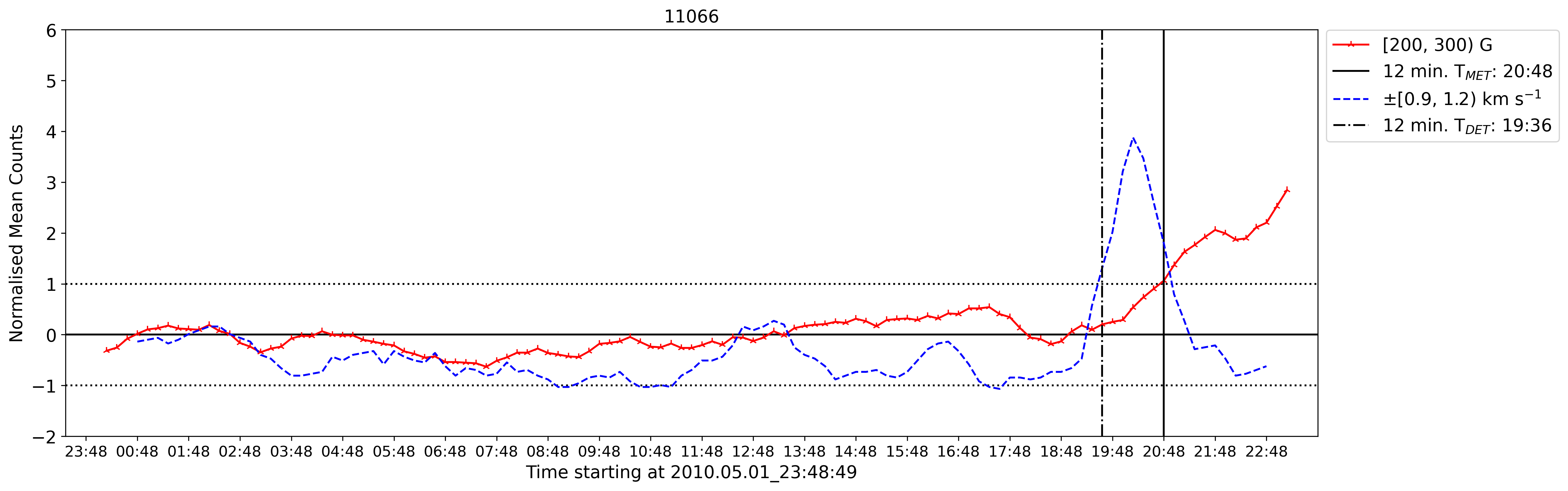}
    \includegraphics[width=\textwidth]{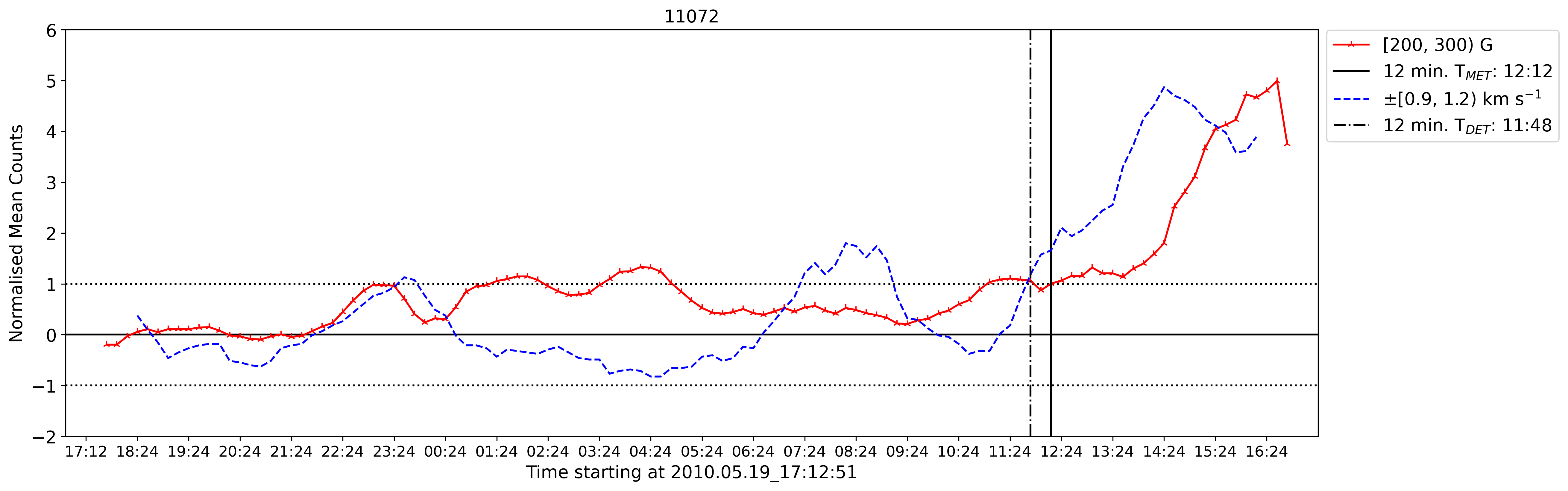}
    \includegraphics[width=\textwidth]{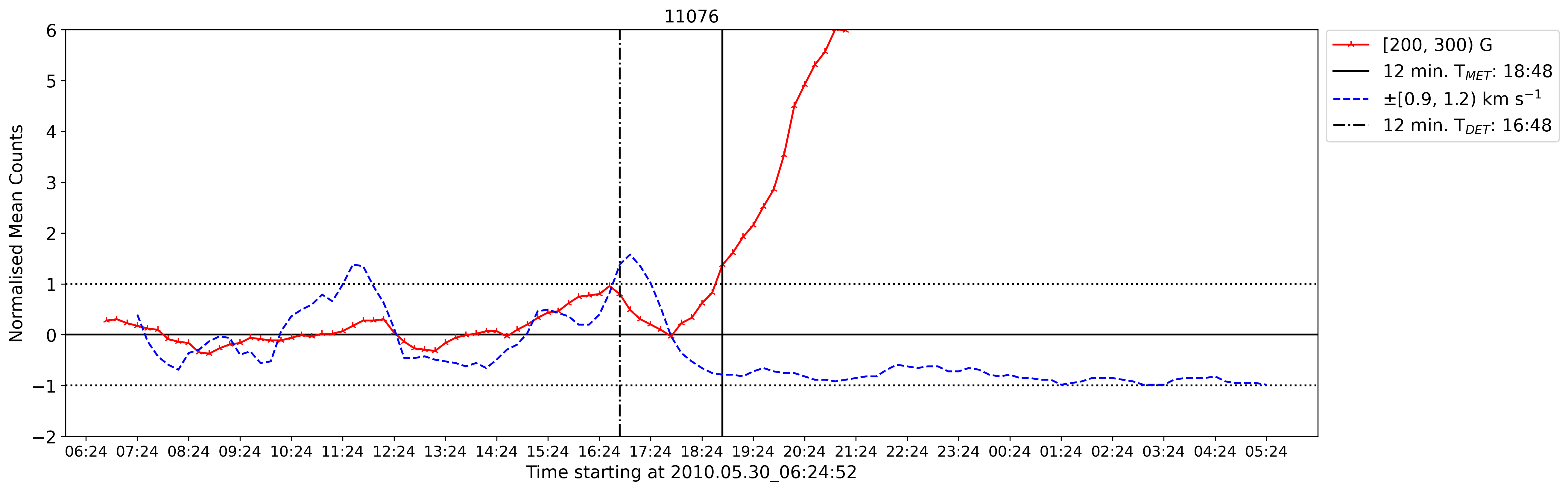}
    \caption{As Fig.~\ref{fig:com_full_res}, but for the 12-minute cadence data. The general behaviour of the ARs is consistent with the full-cadence data with only minor changes to the results returned.}
    \label{fig:com_low_res}
\end{figure*}

Given the inherent errors in the definition of the flux emergence time, we also investigated the $T_{FE}$ and $T_{DET}$ values calculated from lower cadence data in order to understand whether HDF signatures could be detected. This acts as a parameter study aimed at establishing whether this method is scalable to lower-cadence data. 
In Fig.~\ref{fig:com_low_res}, we plot the equivalent to Fig.~\ref{fig:com_full_res} but for the lower $12$ minute cadence.
Similar behaviour is observed between the high-cadence and low-cadence plots for each of these three ARs.
For $T_{FE}$, we now find values of $20$:$48$ UT, $12$:$12$ UT, and $18$:$48$ UT for ARs 11066, 11072, and 11076, respectively. 
These values are also within the same `error' range as the full-cadence data studied previously.
For $T_{DET}$, we find $19$:$36$ UT, $11$:$48$ UT, and $16$:$48$ UT for the three ARs. 
The main difference between these results and the high-cadence results is the return of a potential HDF signature for AR $11076$. 
We note that this is only a small bump in the velocity bin window which falls back below the $1\sigma$ level after three frames, indicating it may not be a true HDF.

An initial comparison may suggest that the differences in the results for $T_{FE}$ between the high- and low-cadence are quite large (several hours), however, close examination of image sequences shows that both are returning results close to the early (ill-defined) on-set of emergence.
There are also several unrelated processes occurring within the FOVs which may explain the small differences in times reported.
In the image sequences of AR 11066, we see the fragmentation and cancellation of a decaying bipole, however, this occurs at the edge of the FOV leaving some flux just outside the FOV during the three hours used for referencing. 
Thus, an increase in the weakest field strength bin and a co-temporal decrease in the other bins is perhaps to be expected as the flux decays, however, this will effect the standard deviations returned differently depending on the frames sampled.
This flux does not appear to be related to the emerging flux, however, its presence in the FOV could explain small (several hour) differences in the emergence time reported here. 
For AR 11072, the background time contains some already emerging weak flux which appears to separate and grow during this time. 
This means that we are potentially seeing emergence signatures in our background at the $1\sigma$ level. 

Overall, it is clear that numerous effects will influence the values returned for $T_{FE}$ and $T_{DET}$, including but not limited to pre-processing steps applied to the data, the time-period selected as the background, and the properties (e.g. cadence) of the data studied. 
Given these issues, it is unclear whether these increases in the number of pixels within specific velocity windows, known as the HDF, can be used as a predictive tool on its own for defining when flux will emerge routinely for all ARs, especially when it is searched for using a large FOV (as is done both here and in \citealt{2014ApJ...794...19T}).
However, the presence of an HDF combined with an increase in flux (within several hours) may indicate that the emerging field may develop further into a full blown AR in the future. 
It is, therefore, still important to study whether this signature can be detected in different velocity bins. 
Due to the similarity in results between the high-cadence and low-cadence analyses, we use 12-minute cadence data throughout the rest of this study when searching for HDFs across a number of different velocity bins.

\subsection{Statistical Analysis}\label{sec:mag_res}

\begin{figure*}[ht!]
    \centering
    \includegraphics[width=\textwidth]{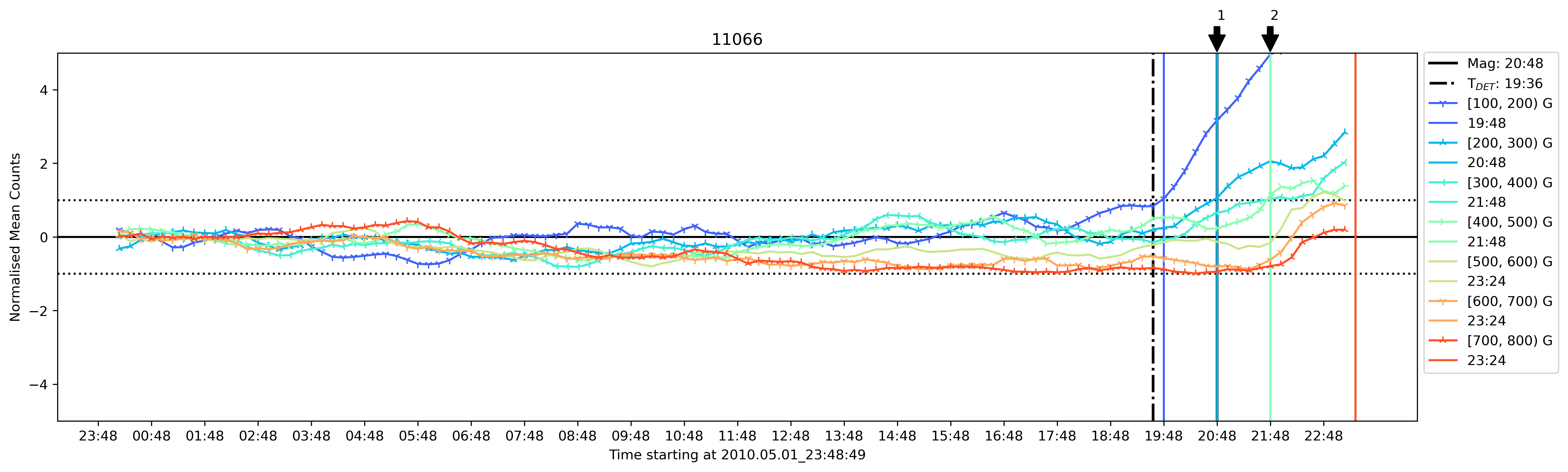}
    \includegraphics[width=\textwidth]{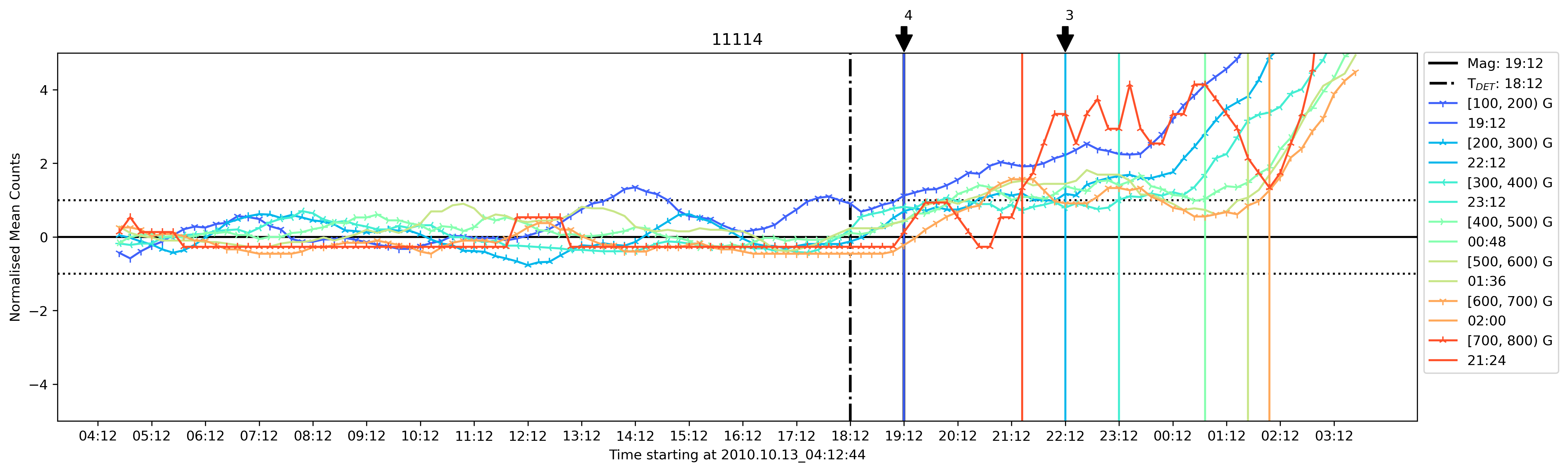}
    \includegraphics[width=\textwidth]{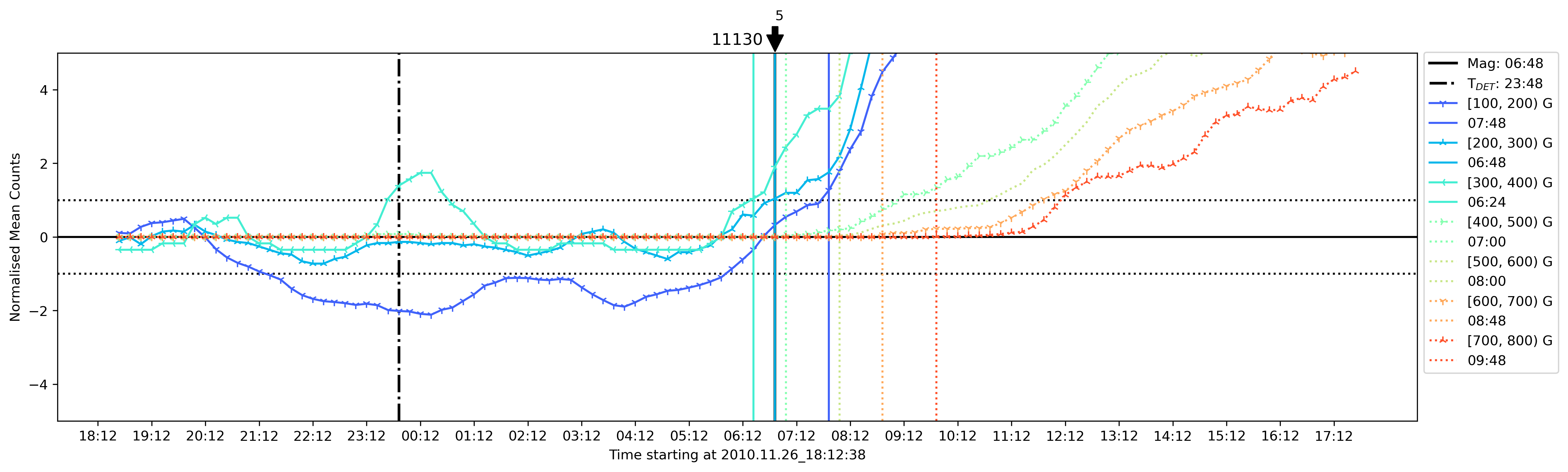}
    \caption{Magnetogram profiles for three representative ARs. Each panel contains seven lines, individually coloured as denoted in the legend, plotting magnetic field windows of 100 G width over the range 100-800 G. All profiles have been normalised against the $1\sigma$ value for that specific magnetic field window calculated individually for each AR. Solid lines denote magnetic field windows that have been normalised against their reference $\sigma$, whilst dotted lines denote those for which that was not possible. Horizontal black lines have been placed at y=0 and $\pm$1 for reference. The vertical coloured lines show the time at which the window either exceeds 1$\sigma$ or, in cases where $\sigma$ could not be calculated, where it is zero for the last time. Arrows are ordered by their place of reference within the text. The complete figure set (16 images) is available in the online journal.}
    \label{fig:sig_mag_profiles_unsmo}
\end{figure*}

\subsubsection{Magnetic Field Data}

In order to further progress towards the aims of this work, we now conduct a statistical analysis of the thresholding method applied to all $16$ ARs using the $\pm$[0.9, 1.2) km s$^{-1}$ velocity window using the 12-minute cadence data. Initially, we study the magnetogram profiles to determine the time of magnetic flux emergence, $T_{FE}$, for each AR. The results of this analysis are presented in the second column of Table~\ref{tab:ar_results}. For a better comparison with \citet{2012ApJ...751..154T}, we report the [200, 300) G magnetic field strength window unless otherwise specified. In the case that no threshold ($1\sigma$) was calculable due to the counts in any specific magnetic field strength window being 0 across the reference time range, we instead find the last time the profile is at 0 counts. This was usually only necessary for sets above 500~G and, as such, has no significant influence on the results presented in Table~\ref{tab:ar_results}.

\begin{figure*}[ht!]
    \centering
    \includegraphics[width=\textwidth]{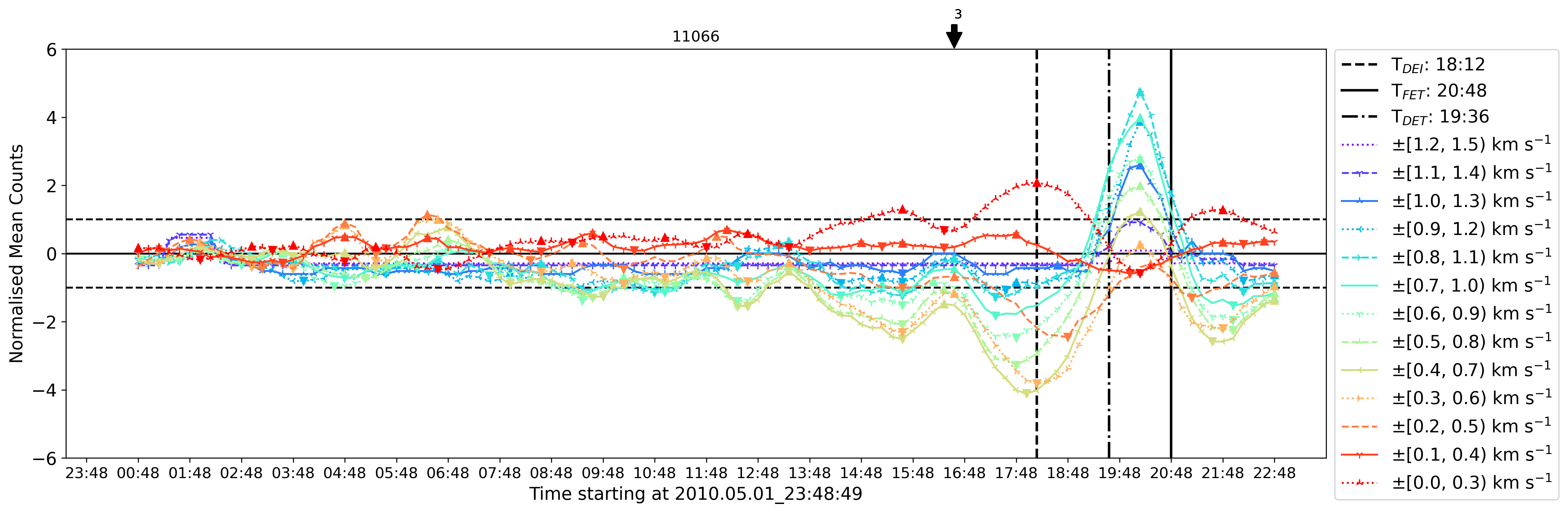}
    \includegraphics[width=\textwidth]{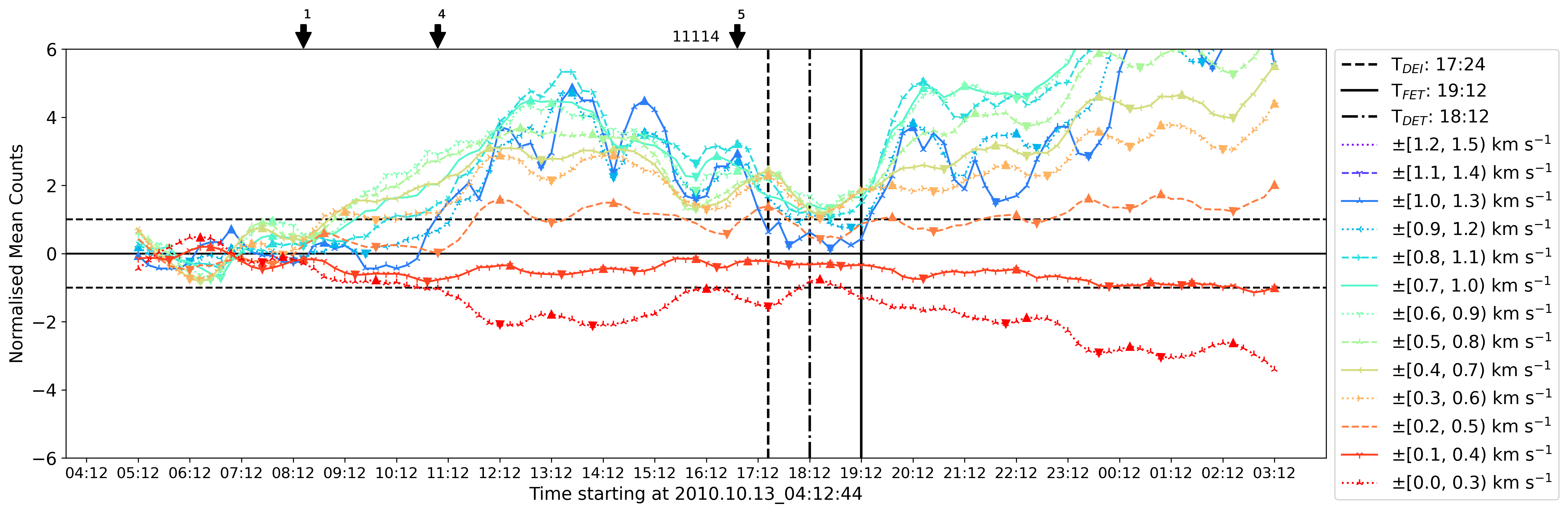}
    \includegraphics[width=\textwidth]{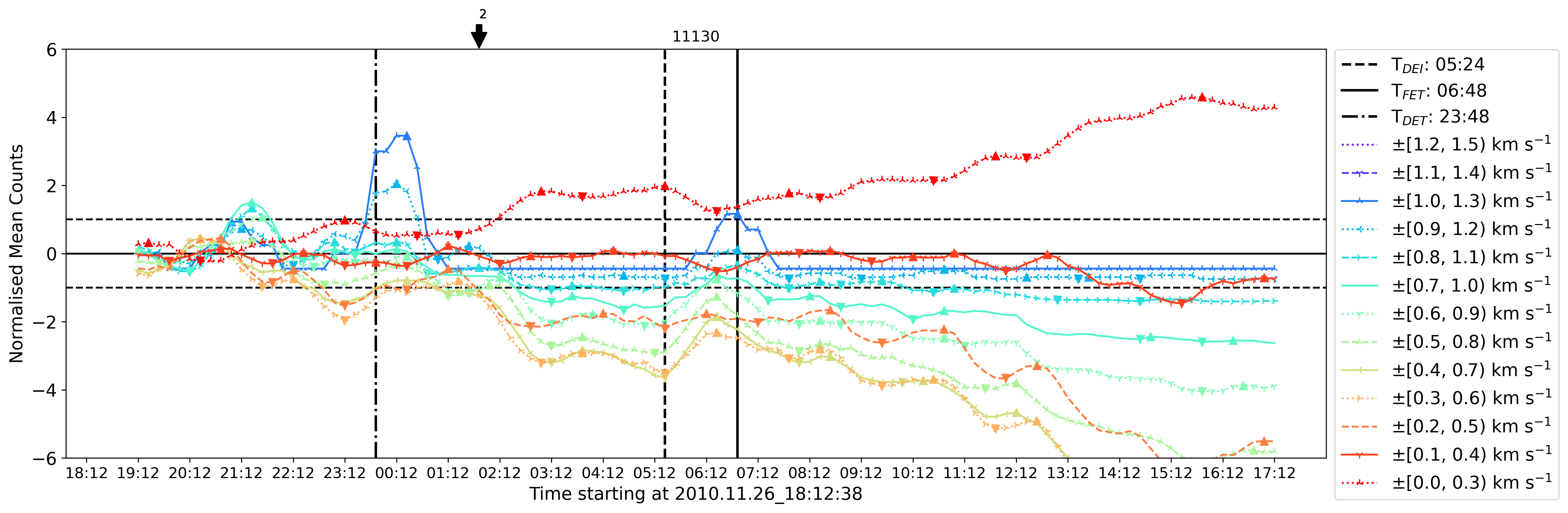}
    \caption{Same as for Fig.~\ref{fig:sig_mag_profiles_unsmo} but for the corresponding smoothed dopplergram profiles. Each panel plots $13$ different Doppler velocity windows, each covering 6 velocity bins of width $0.1$ km s$^{-1}$ over the range $|0|-|1.5|$~km~s$^{-1}$. Each Doppler velocity window is individually coded by colour, line style, and marker as labelled in the legend. The solid black vertical line denotes the time of magnetic emergence, the vertical dot-dashed black line indicates the time of pre-emergence identified using the thresholding method, and the dashed black vertical line locates the determined time of pre-emergence estimated using the inflection point method. Arrows are ordered by place of reference within text. The complete figure set (16 images) is available in the online journal.}
    \label{fig:sig_dop_profiles_smo}
\end{figure*}

In Fig.~\ref{fig:sig_mag_profiles_unsmo}, we present the time profiles of sets of absolute magnetic flux for three example ARs (namely NOAA ARs 11066, 11114, and 11130) for each of the seven magnetic field strength windows.
These plots help detail how we select an appropriate $T_{FE}$ for each AR.
In these plots, solid lines denote that the magnetic field window was normalised against $\sigma$, and dotted lines denote that this was not possible.
The sets have been colour-coded from dark blue ([100, 200) G), through green ([400, 500) G), to red ([700, 800) G), with the specific colours noted in the legend.
The coloured vertical lines denote the time at which a set either exceeded the threshold, or stopped being 0. 
If the set did neither, no vertical line is drawn for that set. 
Note that a vertical line may not seem to appear if another is drawn over it (i.e at the same time), but the time is still reported in the legend.
In the top panel of Fig.~\ref{fig:sig_mag_profiles_unsmo}, we show the time profiles of AR 11066.
Here, the time of emergence is visually clear, and starts around $20$:$48$ UT in the [200, 300) G ( see mid blue vertical line with arrow ``$1$'' above) set, with the [300, 400) G, and [400, 500) G sets  see light green vertical line with arrow ``$2$'' above) both finding a time of $21$:$48$ UT. 
We can also see that a few profiles ($>600~$G) did not exceed the threshold over the course of the analysed time period, though the [500, 600) G did exceed $1\sigma$ briefly before returning under. 
In the middle panel of Fig.~\ref{fig:sig_mag_profiles_unsmo}, we present the magnetic profiles of AR 11114. 
Here, we see emergence begins at $22$:$12$ UT in the [200, 300) G set  (see arrow ``$3$''), with the majority of remaining profiles crossing the threshold over the following hours. 
This event is of interest here as, when the image sequences of the event are studied, it becomes evident that the [100, 200) G profile  (see arrow ``$4$'') in this case is more appropriate. 
This is due to the emergence having evolved significantly by 22:12 UT  (arrow ``$3$'') when the higher magnetic field strength window begins to display emergence signatures.
Finally, in the bottom panel of Fig.~\ref{fig:sig_mag_profiles_unsmo}, we plot the time-profiles of AR 11130.
Here, we immediately see that there is little to no activity over 400~G pre-emergence, with the majority of activity located in the lowest three magnetic field strength windows.
Nonetheless, we see general agreement between these three windows and we, therefore, consider the [200, 300) G window giving 06:48 UT  see arrow ``$5$'') as the time of emergence.
Interestingly, we see greater difference between the higher and lower profiles than in the other ARs in this panel, though the reasons are not investigated here. 
Comparing the profiles of these three ARs we see the common trend of a strong, possibly non-linear increase (as would be expected from flux emergence, though we do not test non-linearity here), across multiple sets of magnetic field strengths over the course of several hours, supporting our assertion that differences of several hours in $T_{FE}$ are acceptable.

\subsubsection{The $[0.9, 1.2)$ $km$ $s^{-1}$ Doppler Velocity Window} \label{sec:thres}

In Fig.~\ref{fig:sig_dop_profiles_smo}, we plot the equivalent to Fig.~\ref{fig:sig_mag_profiles_unsmo} for these three ARs except for the Doppler velocity bins.
Likewise, these bins have been colour-coded from dark blue ($\pm$[1.2, 1.5) km~s$^{-1}$), through green ($\pm$[0.5, 0.8) km~s$^{-1}$), to red ($\pm$[0.0, 0.3) ~km~s$^{-1}$).
Furthermore, each set has been given a unique combination of line style and marker to help distinguish them. 
We begin our analysis by studying the $\pm$[0.9, 1.2) km s$^{-1}$ window in order to complement Sect.~\ref{sec:tor_com}.
The solid black vertical line represents $T_{FE}$ and the dot-dashed vertical line represents the HDF signature, $T_{DET}$, from the $\pm$[0.9, 1.2) km s$^{-1}$ window.  
For AR 11066 (top panel), the clear rise in the Doppler windows sampling higher velocities is immediately clear at $19$:$36$  (see black dot-dash line) UT representing the HDF. 
For AR 11114 (middle panel), the Doppler velocity windows are much more complex, with large increases in the number of pixels detected in numerous windows rising from around $08$:$00$ UT  (see below arrow ``$1$''), before falling below the $1\sigma$ value again at around $18$:$00$ UT  (see between black dash and dot-dash vertical lines). 
This is followed by a subsequent rise in the [0.9, 1.2) km s$^{-1}$ velocity window at $18$:$12$ UT  (see black dot-dash vertical line). 
This time is, therefore, recorded as the HDF. 
For AR 11130 (bottom panel), we visually represent one of the potential problems with the thresholding method. 
This is that increases well before the time of emergence (more than $6$ hours) can be picked up and returned by an automated code. 
These can be both transient such as for the higher Doppler velocity windows which peak around $00$:$00$ UT  (see black dot-dash vertical line), or more sustained such as for the [0.0, 0.3) km s$^{-1}$ Doppler velocity window  (see red dotted profile [topmost] with asterisks, below arrow ``$2$'').
Either way, it is difficult to return a clear $T_{DET}$ for this AR.

In the third column of Table~\ref{tab:ar_results}, we present the results of the thresholding method for the [0.9, 1.2) km s$^{-1}$ Doppler velocity window.
We find confident HDF signatures for $6$ of the $16$ ARs studied here. 
The thresholding technique returns results for $5$ further ARs, however, the HDF signature is found to preceed the flux emergence by more than $7$ hours in each case (indicated by an `f' in the fourth column of Table~\ref{tab:ar_results}).
We, therefore, ignore those ARs in this analysis. 
From the $6$ ARs that we return confident HDF signatures from, we find lead times, ${\Delta}T$, of between $12$ and $120$ minutes (see the fourth column of Table~\ref{tab:ar_results}), with an average of $58$ minutes.
Both the percentage ($37.5$\%) of ARs displaying HDF signatures and the lead times returned by the thresholding method ($58$ minutes) match well with the results of \citet{2014ApJ...794...19T} ($26$\% and $61$ minutes, respectively) giving us confidence that the lower cadence data studied here does not significantly influence the results.
Equivalent plots to Figs.~\ref{fig:sig_mag_profiles_unsmo} and \ref{fig:sig_dop_profiles_smo} are included for each analysed AR in the supplementary material for completeness. 

\begin{deluxetable}{ccccccc}[ht]
\tablecaption{Magnetic flux emergence time and pre-emergence signature times (identified using two methods) for each of the $16$ ARs studied here. \label{tab:ar_results}}
\tablewidth{0pt}
\tablehead{
\colhead{NOAA} &\colhead{$\hat{B}$}&\colhead{V$_{Dop}$}&\colhead{$\Delta T$}&\colhead{V$_{Dop}$}&\colhead{$\Delta T$} \\
\colhead{\#} &\colhead{T$_{FE}$}&\colhead{T$_{DET}$}\tablenotemark{a}  & \colhead{$\Delta T_T$}\tablenotemark{a}&\colhead{T$_{DEI}$}\tablenotemark{b}& \colhead{$\Delta T_I$}\tablenotemark{b} 
}
\startdata
11066 & 20:48 & 19:36 & 72 & 18:12 & 156 \\
11072 & 12:12 & 11:48 & 24 & 11:12 & 60 \\
11076 & 18:48 & 16:48 & 120 & 17:36\tablenotemark{g} & 72 \\
11088 & 04:48\tablenotemark{c} & 04:36 & 12 & 02:48\tablenotemark{g} & 120 \\
11114 & 19:12\tablenotemark{d} & 18:12 & 60 & 17:24 & 108 \\
11116 & 12:12 & -- & -- & 10:36\tablenotemark{g} & 96 \\
11122 & 11:00 & 03:48 & 432\tablenotemark{f} & -- & -- \\
11130 & 06:48 & 23:48 & 420\tablenotemark{f} & 05:24 & 84 \\
11136 & 18:00 & -- & -- & -- & -- \\
11142 & 03:24 & 14:24 & 780\tablenotemark{f} & 01:48 & 96 \\
11148 & 13:12 & 06:00 & 432\tablenotemark{f} & -- & -- \\
11294 & 08:24 & 07:24 & 60 & -- & -- \\
11400 & 16:12\tablenotemark{c} & -- & -- & 14:36\tablenotemark{g} & 96 \\
11414 & 01:00\tablenotemark{e} & -- & -- & -- & -- \\
11472 & 19:48 & -- & -- & -- & -- \\
11523 & 13:36 & 05:24 & 492\tablenotemark{f} & 11:12 & 120 \\
\enddata

\tablenotetext{a}{Thresholding method applied to the [0.9, 1.2) km s$^{-1}$ window using $12$ minute cadence data (see Sect.~\ref{sec:thres}).}
\tablenotetext{b}{Inflection point method using $12$  minute cadence data (see Sect.~\ref{sec:infle}).}
\tablenotetext{c}{Taken from [300,~400~G).}
\tablenotetext{d}{Taken from [100,~200~G).}
\tablenotetext{e}{Taken from [600,~700~G).}
\tablenotetext{f}{Not deemed confident HDF.}
\tablenotetext{g}{Tentative signature.}

\end{deluxetable}

\subsubsection{Other Doppler Velocity Windows}

\begin{deluxetable*}{ccccccccccccccccc}[ht]
\rotate
\centering
\tablecaption{Time of HDF detection, $T_{DET}$, for each studied velocity window for the 16 ARs analysed here. The column labels indicate the specific velocity window in km~s~$^{-1}$ ($\pm$). The results are split into four classifications, namely: `0' where the absolute value of the profile never increases above $1\sigma$; `1' where the absolute value of the profile goes above $1\sigma$ but returns below this level before $T_{FE}$; `2' where $T_{FE}$ and $T_{DET}$ are either separated by one time-step or are equal; and `3' where the absolute value of the profile increases above $1\sigma$ and remains above that value until after $T_{FE}$. Positive and negative values of 1, 2, and 3 indicate whether the velocity window goes above $1\sigma$ or below $-1\sigma$, respectively.  \label{tab:ar_profile_type}}
\tablewidth{0pt}
\tablehead{
\colhead{AR \#} & \colhead{[1.2-1.5)} & \colhead{[1.1-1.4)} & \colhead{[1.0-1.3)} & \colhead{[0.9-1.2)} & \colhead{[0.8-1.1)} & \colhead{[0.7-1.0)} & \colhead{[0.6-0.9)} &\colhead{[0.5-0.8)}  & \colhead{[0.4-0.7)} & \colhead{[0.3-0.6)} &\colhead{[0.2-0.5)} & \colhead{[0.1-0.4)} & \colhead{[0.0-0.3)}}
\startdata
11066 & 0: -- & 0: -- & 1: 19:48\tablenotemark{a} & 3: 19:36\tablenotemark{a} & 3: 19:24\tablenotemark{a} & 3: 19:24\tablenotemark{a} & 1: 19:36\tablenotemark{a} & 1: 19:48\tablenotemark{a} & 1: 20:00\tablenotemark{a} & -1: 14:00 & -1: 17:24 & 0: -- & 1: 17:00 \\ 
11072 & 0: -- & 2: 12:12\tablenotemark{a} & 2: 12:00\tablenotemark{a} & 3: 11:48\tablenotemark{a} & 3: 11:48\tablenotemark{a} & 1: 08:00 & 1: 08:24 & 1: 08:36 & -1: 11:24\tablenotemark{a} & -3: 09:36 & -2: 12:12\tablenotemark{a} & 0: -- & -1: 06:48 \\ 
11076 & 0: -- & 0: -- & 1: 16:36 & 1: 16:48\tablenotemark{a} & -2: 18:36\tablenotemark{a} & -3: 18:00\tablenotemark{a} & -3: 17:24\tablenotemark{a} & -3: 16:36 & -3: 12:12 & -3: 12:00 & -3: 16:24 & 0: -- & 3: 16:24 \\ 
11088 & 1: 23:36 & 2: 04:36\tablenotemark{a} & 2: 04:36\tablenotemark{a} & 2: 04:36\tablenotemark{a} & 2: 04:48\tablenotemark{a} & 1: 18:24 & 1: 18:24 & 1: 18:12 & 1: 17:48 & 1: 18:12 & 1: 21:36 & -1: 19:24 & -1: 02:48\tablenotemark{a} \\ 
11114 & 0: -- & 1: 15:00 & 1: 11:00 & 1: 18:12\tablenotemark{a} & 3: 10:12 & 3: 09:24 & 3: 09:12 & 3: 09:00 & 3: 09:00 & 3: 10:24 & 1: 17:00 & 0: -- & -2: 19:00\tablenotemark{a} \\ 
11116 & 0: -- & 0: -- & 0: -- & 0: -- & 1: 10:24\tablenotemark{a} & 0: -- & -1: 09:24 & -1: 11:12\tablenotemark{a} & -1: 11:24\tablenotemark{a} & -1: 09:00 & -1: 09:12 & 1: 02:00 & 0: -- \\ 
11122 & 0: -- & 0: -- & 1: 03:48 & 1: 03:48 & 0: -- & -3: 09:48\tablenotemark{a} & -3: 09:48\tablenotemark{a} & -3: 09:00\tablenotemark{a} & -3: 08:12 & -3: 07:36 & -3: 09:00\tablenotemark{a} & 0: -- & 3: 10:24\tablenotemark{a} \\ 
11130 & 0: -- & 0: -- & 2: 06:36\tablenotemark{a} & 1: 23:48 & -1: 04:12 & -1: 02:36 & -2: 06:36\tablenotemark{a} & -3: 02:24 & -3: 01:00 & -3: 01:48 & -3: 01:48 & 0: -- & 3: 02:12 \\ 
11136 & 0: -- & 0: -- & 1: 15:00 & 0: -- & 0: -- & 1: 17:24\tablenotemark{a} & 3: 17:12\tablenotemark{a} & 3: 17:00\tablenotemark{a} & -1: 14:12 & -1: 14:12 & -1: 14:24 & 0: -- & 0: -- \\ 
11142 & 0: -- & 1: 00:24 & 1: 14:36 & 1: 14:24 & 1: 00:36 & 1: 00:36 & 2: 03:12\tablenotemark{a} & 3: 02:48\tablenotemark{a} & 3: 02:36\tablenotemark{a} & 3: 02:48\tablenotemark{a} & 3: 03:00\tablenotemark{a} & -1: 20:48 & -3: 03:00\tablenotemark{a} \\ 
11148 & 0: -- & 0: -- & 1: 10:36 & 1: 06:00 & 1: 06:12 & 0: -- & 0: -- & 0: -- & 0: -- & 0: -- & 0: -- & 0: -- & 0: --  \\ 
11294 & 0: -- & 0: -- & 3: 07:24\tablenotemark{a} & 3: 07:24\tablenotemark{a} & 3: 07:24\tablenotemark{a} & 3: 07:24\tablenotemark{a} & 3: 07:24\tablenotemark{a} & 3: 07:24\tablenotemark{a} & 3: 08:00\tablenotemark{a} & 0: -- & 0: -- & 0: -- & -2: 08:24\tablenotemark{a} \\ 
11400 & 0: -- & 0: -- & 0: -- & 0: -- & -2: 16:00\tablenotemark{a} & -3: 15:48\tablenotemark{a} & -3: 13:12 & -3: 10:00 & -3: 10:00 & -3: 10:00 & -3: 10:12 & 0: -- & 3: 13:00 \\ 
11414 & 0: -- & 0: -- & 0: -- & 0: -- & 0: -- & 0: -- & 0: -- & 1: 16:12 & -2: 00:48\tablenotemark{a} & -3: 00:24\tablenotemark{a} & -3: 00:24\tablenotemark{a} & 0: -- & 0: -- \\ 
11472 & 0: -- & 1: 19:36\tablenotemark{a} & 0: -- & 0: -- & 1: 17:48\tablenotemark{a} & 1: 19:00\tablenotemark{a} & 1: 19:24\tablenotemark{a} & 1: 18:00\tablenotemark{a} & -1: 09:24 & -1: 14:36 & 0: -- & 0: -- & 0: --  \\ 
11523 & 2: 13:12\tablenotemark{a} & 3: 09:12 & 3: 08:24 & 3: 05:24 & 3: 05:12 & 3: 05:36 & 3: 05:48 & 3: 06:00 & 3: 07:00 & 3: 09:12 & 3: 12:24\tablenotemark{a} & -3: 10:12 & -3: 07:24 \\ 
\enddata

\tablenotetext{a}{Within two hours of $T_{FE}$.}

\end{deluxetable*}

We now attempt to calculate $T_{DET}$ using the thresholding method for a larger number of Doppler velocity windows. The results from this analysis are presented in Table~\ref{tab:ar_profile_type}. Each result is split into one of four categories depending on whether/how the profile increases over $1\sigma$. We do not differentiate between increases over $1\sigma$ and decreases below $1\sigma$ meaning these categories can be either positive (increases over $1\sigma$) or negative (decreases below $1\sigma$). Category 0 indicates that the absolute value of the Doppler velocity window never increased above $1\sigma$. Category 1 indicates that the absolute value of the profile increased over $1\sigma$ at some point during the analysed time, but that it had decreased back below the $1\sigma$ value before $T_{FE}$. Category 2 indicates that the window increased over $1\sigma$ within one frame of $T_{FE}$. Finally, category 3 indicates that the Doppler velocity window increased over $1\sigma$ and stayed above that value until after $T_{FE}$.

It is immediately evident that different Doppler velocity windows return different results, with some appearing more effective in detecting HDFs than others. Notably, the $\pm$[1.2, 1.5) km s$^{-1}$ and $\pm$[0.1, 0.4) km s$^{-1}$ Doppler velocity windows return very few potential results. The superscript `a' in Table~\ref{tab:ar_profile_type} indicates whether a profile crosses the $1\sigma$ threshold within two hours of $T_{FE}$ for any given AR. We consider two hours to be an appropriate upper limit for our discussion as it is double the average HDF lead-time reported in the previous subsection and in \citet{2014ApJ...794...19T}. Given that the method of \citet{2014ApJ...794...19T} is limited to increases over a positive threshold value, we will initially focus only on positive potential HDF signatures within two hours of $T_{FE}$. In this sense, the $\pm$[0.9, 1.2) km s$^{-1}$ and $\pm$[0.8, 1.1) km s$^{-1}$ Doppler velocity windows return the most results, with both finding $6$ potential HDF signatures. Several other Doppler velocity windows return $4$ or $5$ potential HDF signatures within two hours of $T_{FE}$, with only the $\pm$[0.1, 0.4) km s$^{-1}$ window returning no results. Considering the values with negative significance (i.e. the profiles which drop below $-1\sigma$ within two hours of $T_{FE}$), we find a different picture with the $\pm$[0.0, 0.3) km s$^{-1}$ Doppler velocity window identifying the most potential HDF signatures with $4$. No negative potential HDF signatures (at any time before $T_{FE}$) were identified in the Doppler velocity windows included in the first four columns of Table~\ref{tab:ar_profile_type}, suggesting analysis of negative values may only provide results at lower Doppler velocities. 

Our analysis has shown that if we consider a wider array of Doppler velocity windows, we are able to increase the number of positive potential HDF signatures within two hours of $T_{FE}$. From the ARs studied, we are able to move from $6$ (out of $16$) ARs from which HDFs can be identified from the $\pm$[0.9, 1.2) km s$^{-1}$ Doppler velocity window to $13$ ARs from which HDFs can be detected when considering all Doppler velocity windows. One notable example of an AR from which a potential signal can now be detected is AR $11122$, which displays positive significance only in the $\pm$[0.0, 0.3) km s$^{-1}$ Doppler velocity window. This AR emerged close to disk centre (at a longitude of around $-25^\circ$) potentially explaining why the clear HDF signatures detected in higher Doppler velocity windows at larger $\mu$-angles may not be visible. 
Instead the rising flux may inhibit local convection leading to an increase in pixels which display very small Doppler velocities, as may be seen in \citet{2007A&A...467..703C}. 

\begin{figure*}[ht!]
    \centering
    \includegraphics[width=0.49\textwidth]{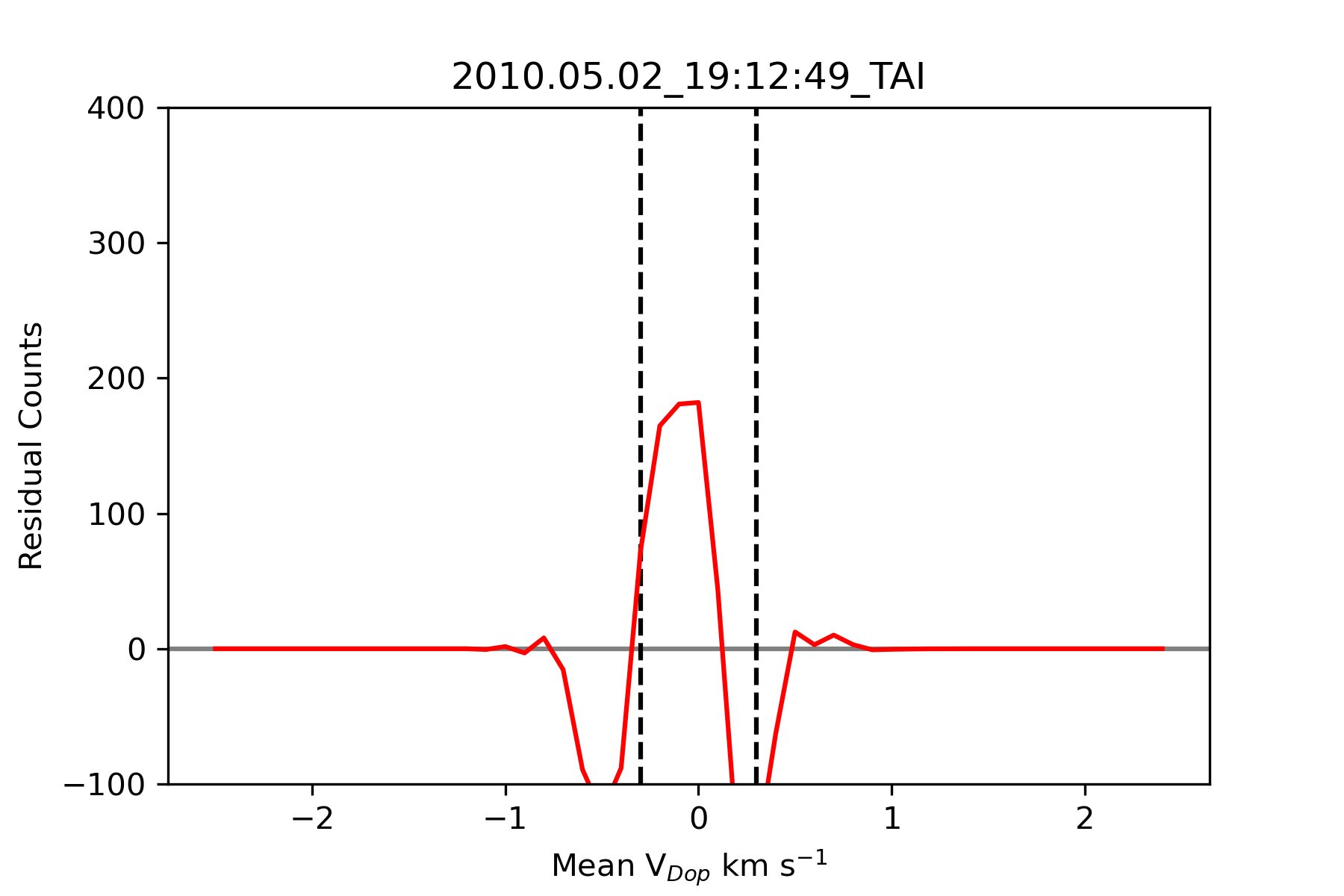}
    \includegraphics[width=0.49\textwidth]{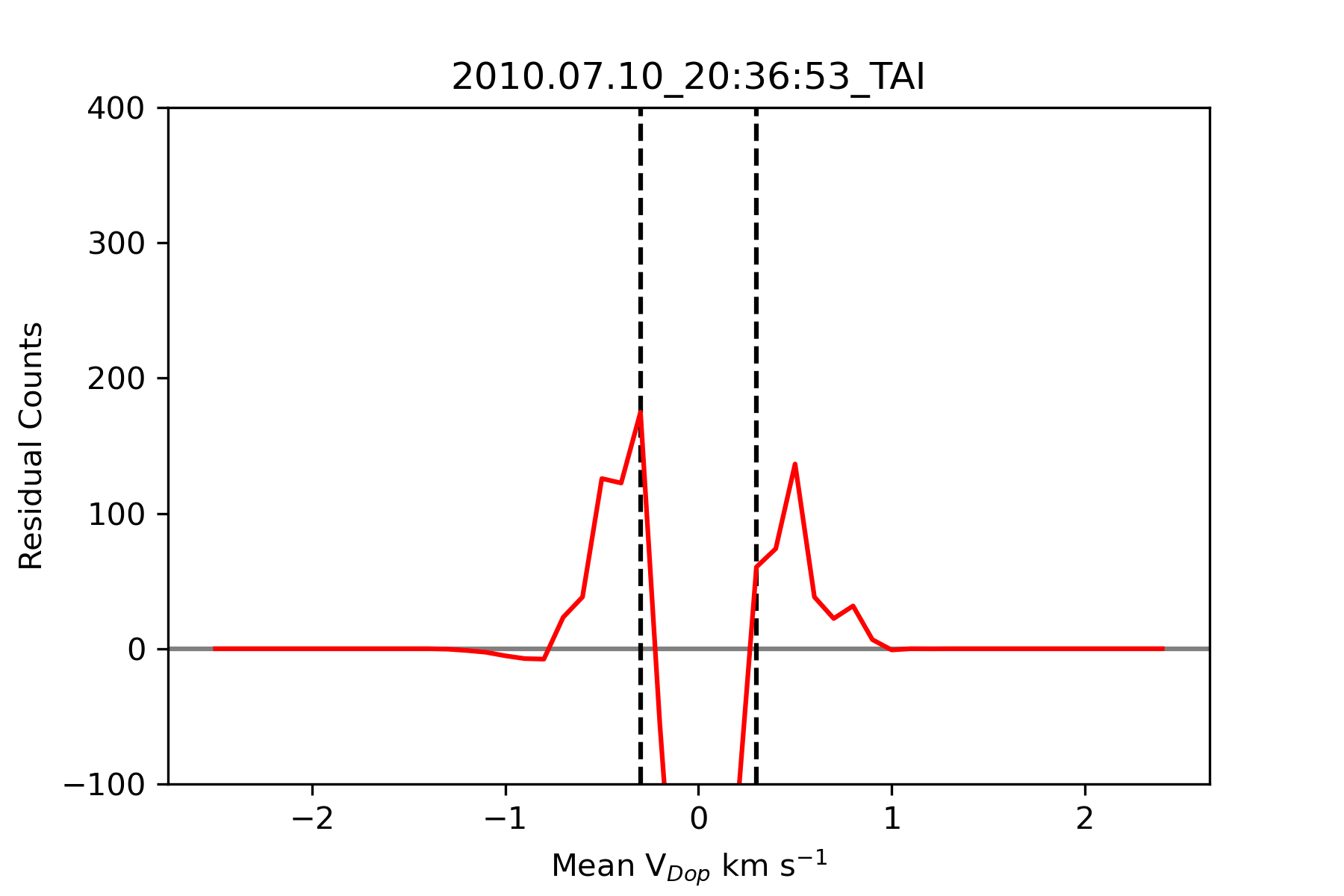}
    
    \includegraphics[trim=3.5cm 0cm 2.5cm 0cm,
    width=\textwidth]{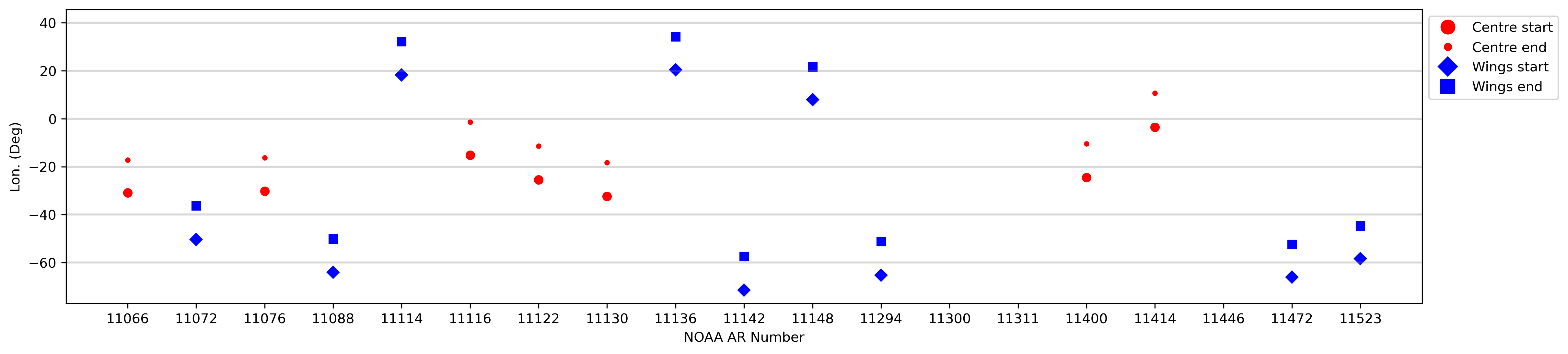}
    \caption{ Top row:  Panel a) The mean residual counts across the velocity profile for AR 11066 at the shown time showing a ``centre'' increase. 
    Panel b) As a) for AR 11088 showing a ``wings'' increase.
    In both a) and b) the vertical lines denote the wing/centre limit, $\pm$0.3~km~s$^{-1}$. 
    Panel c) The longitude versus AR. Colour and shape are coded by trend type, where red circles show centre trends, and blue squares  and diamonds show wing trends. The grey lines are to help guide the reader's eye.
    }
    \label{fig:lon_vs_ar}
\end{figure*}

\subsection{Collective Group Behaviour}\label{sec:infle}

Following on from our statistical analysis, we now try to develop a technique for detecting pre-emergence which does not rely on one single Doppler velocity window, but instead makes use of the collective group behaviour across multiple windows. It is here that our proposed method of determining the time of pre-emergence diverges significantly from that of \citet{2012ApJ...751..154T} (compare Fig~\ref{fig:mag_dop_flowcharts} M5-8 to D5-8). In this subsection, we instead look for transient disruptions across a group of Doppler velocity bin profiles, marked by collective changes in relative amplitudes. 
We specifically look for times when multiple Doppler velocity windows peak (or trough) identified by `inflection points'.
This allows us to search profiles that have higher thresholds throughout their evolution (such as AR 11130) and profiles that do not match the implicit ``below-until-event'' assumption made in \citet{2012ApJ...751..154T}.
This presents the problem of what feature within the profile to choose, when each is inherently noisy and oscillatory.
Due to this, we choose to mark our proposed time of pre-emergence as the final inflection point before magnetic emergence, as this is unambiguous and replicable. 
Currently, the selection of which group of inflection points to consider is done manually in conjunction with careful analysis of image sequences, with automated methods in development.

We will now discuss the inflection point method by using the three ARs plotted in Fig.~\ref{fig:sig_dop_profiles_smo} as examples.
The times of the potential pre-emergence signatures for this method, $T_{DEI}$, are denoted by the vertical dashed black lines. 
For AR 11066 (top panel), at around 16:24  (see arrow ``$3$'') the Doppler profiles clearly start to show a sudden and rapid change to the amplitude of the signal, reaching a peak (or inflection point) at 18:12 UT  (see black dashed vertical line). 
This is both 156 minutes prior to the magnetic emergence at 20:48 UT [9:] (black solid vertical line) and $70$ minutes prior to the HDF time reported for this AR by \citet{2014ApJ...794...19T}.
We see the inflection point in two forms. The first is the decrease and convergence of a number of higher velocity sets ($\geq |0.9|$~km~s$^{-1}$), and the second is the increase and convergence of a number of lower velocity sets, ($\leq0.8$~km~s$^{-1}$)  (see arrow ``$3$'').
For AR 11114 (middle panel), numerous profiles increase above (decrease below) the $1\sigma$ ($-1\sigma$) value at around $11$:00 UT  (see arrow ``$4$''), eight hours before $T_{FE}$.
A number of the profiles remain above $|1\sigma|$ for the entire time, however, a clear example of collective group behaviour is apparent starting at around $16$:$30$ UT  (see arrow ``$5$'').
This results in an inflection point at $17$:$24$ UT, $108$ minutes prior to $T_{FE}$  (see dashed black vertical line).
For AR 11130 (bottom panel), an increase above the $1\sigma$ threshold occurs at around 01:36 UT  (see arrow ``$2$'') for the [0.0, 0.3) km s$^{-1}$ Doppler velocity window which is too early to be associated with an HDF.
The inflection point becomes very useful for estimating pre-emergence in this case, identifying a pre-emergence time of 05:24 UT  (see dashed black vertical line) which is 84 minutes before the magnetic emergence at 06:48 UT  (see solid black vertical line).
Here the change in amplitude is seen relative to already diverging profiles, making it more difficult to judge the precise time of change.
We suggest that this inflection point could be due to the sub-photospheric flux disrupting the turbulent motions at the surface, as may also be seen in AR 11114. 

In the fifth and sixth columns of Table.~\ref{tab:ar_results}, we present the time of the inflection point and the difference between this time and $T_{FE}$, denoted by $\Delta T_I$.
We find collective group behaviour resulting in an inflection point for $10$ (out of $16$) ARs, finding an average $\Delta T_I$ of $100.8$ minutes, nearly double the $58$ minutes found using the thresholding method. 
Of these ARs, four display only potential signatures of inflection points, denoted by the superscript `\textit{g}' in Table~\ref{tab:ar_results}; however, we consider these to be tentative here and include them in our sample. 
Whilst there is a recoverable inflection point for most of the studied ARs, there are other features within the Doppler velocity profiles of the other ARs that make it harder to identify.
For the six ARs from which no inflection point is recoverable, this is typically because there is no clear change in amplitude across the profiles beyond the underlying oscillation. 
For AR 11294, it is simply due to the magnetic emergence occurring very near to the start of the data set, making judgements about changes in amplitude impossible. 
It is clear that further work must be conducted to optimise this method but we do find it encouraging for the future.

As a final piece of analysis, we study the relationship between longitude and the Doppler velocity windows which are positive during the inflection point.
Here, we define an AR as a `centre' if profiles below the $\pm$[0.3, 0.6)~km~s$^{-1}$ Doppler velocity window show a positive number of counts over the majority of the profile before the time of magnetic emergence. 
A profile is deemed to be a `wing' profile when we see the opposite. 
To demonstrate this longitudinal dependence upon our results, in Fig.~\ref{fig:lon_vs_ar} we show whether each AR shows an increase in the centre or the wings plotted against longitude.
As can be seen in Fig.~\ref{fig:lon_vs_ar} there is clear correlation between the longitude and the resultant trend in the data, with centre trends being found for events with $lon \lessapprox|30|^{\circ}$, and wing trends found for $lon\gtrapprox|20|^{\circ}$. 
We believe that a small transition region between centre and wing trends will become apparent with a larger sample size due to variations between data sets.
This analysis once again highlights the importance of studying multiple Doppler velocity windows.

\section{Discussion} \label{sec:disc}

In this article, we have used two methods to analyse 16 ARs of the 21 denoted as `P0' in \citet{2016A&A...595A.107S} (detailed in both Fig.~\ref{fig:ar_track_locs} and Table~\ref{tab:ar_list}) in order to determine potential pre-emergence signatures using two methods. 
The first method was the thresholding method (as previously used by \citealt{2012ApJ...751..154T, 2014ApJ...794...19T}) and the second method was through the identification of inflection points. 
Both methods were applied to Doppler velocity and magnetic field windows normalised against their own standard deviations, to allow direct comparison across multiple Doppler velocity windows and multiple ARs.
The thresholding method was applied in order to detect signatures of horizontal flows in the solar photosphere prior to the flux emergence (predicted by \citealt{2010ApJ...720..233C}), now known as  Horizontal Divergent Flows (HDFs) (\citealt{2012ApJ...751..154T}). 
As our data preparation method (presented in Figs.~\ref{fig:flowchart} and \ref{fig:mag_dop_flowcharts}) diverged slightly from that originally used by \citet{2014ApJ...794...19T}, we conducted a comparison between our results calculated using the thresholding method and their results for three ARs which over-lapped in our samples. 
Our results, calculated from the $\pm$[0.9, 1.2) km s$^{-1}$ Doppler velocity window and displayed in Fig.~\ref{fig:com_full_res}, returned values of both  the time of magnetic flux emergence, $T_{FE}$, and time of Doppler emergence using the threshold method, $T_{DET}$,  close to those of  \citet{2014ApJ...794...19T} indicating consistency in our results.

We repeated our analysis on the same three ARs using the $\pm$[0.9, 1.2) km s$^{-1}$ Doppler velocity window and the lower cadence 12-minute data  to investigate whether HDFs could still be detected.
Our results (plotted in Fig.~\ref{fig:com_low_res}) match with the results found in our higher cadence data and in \citet{2014ApJ...794...19T} with only minor differences between the ARs being returned. 
One of the larger differences was the detection of a HDF for AR 11076 when no HDF was detected in the higher cadence data. 
However, the increase above the $1\sigma$ level returned as the HDF signature from the $\pm$[0.9, 1.2) km s$^{-1}$ Doppler velocity window was transient, lasting only three frames suggesting this may not have been a true HDF.
Given the similarities in results, we decided to use the lower cadence $12$-minute cadence data for a larger statistical analysis here.
This analysis, using the $\pm$[0.9, 1.2) km s$^{-1}$ Doppler velocity window, returned HDFs for 6 ($37.5$\%) of ARs with an average  time difference (using the threshold time), $\Delta T_T$, of $58$ minutes. 
These results match well with \citet{2014ApJ...794...19T} who found clear HDFs for $25$\% of ARs with a $\Delta T_T$ of $61$ minutes.
The results from this analysis can be found in Table~\ref{tab:ar_results}. 

Finally regarding the thresholding method, we conducted an analysis of a wider range of Doppler velocity windows spanning from $\pm$[1.2, 1.5) km s$^{-1}$ to $\pm$[0.0, 0.3) km s$^{-1}$ to investigate whether HDFs could be detected more widely. 
This analysis 
allowed us to increase the number of potential HDF detections (within two hours of $T_{FE}$) to $13$ (out of the $16$ ARs investigated). 
We did not calculate a $\Delta T_T$ value from this analysis as our two hour limit, selected to retrieve general results, would have biased our calculations.
Limiting our analysis to positive results initially (i.e. where the Doppler velocity window increased over the $1\sigma$ level), it was found that the $\pm$[0.9, 1.2) km s$^{-1}$ and $\pm$[0.8, 1.1) km s$^{-1}$ were the most effective Doppler velocity windows for detecting HDFs with both finding $6$ potential HDFs. 
Of particular interest was AR 11122 which only displayed evidence of a potential positive HDF signature in the [0.0, 0.3) km s$^{-1}$ Doppler velocity window, highlighting the benefit of studying a wider range of Doppler velocity windows.
Considering negative results across all ARs (i.e. where the Doppler velocity window dropped below the $-1\sigma$ threshold), the $\pm$[0.0, 0.3) km s$^{-1}$ window appeared to display the most (four) potential signatures of the HDF (within two hours of $T_{FE}$). 
The inclusion of negative results could be important when the positive HDF signal is blurred across multiple Doppler velocity windows (e.g. closer to disk centre), potentially meaning it does not increase above the $1\sigma$ value in any. 

Given that potential HDF signatures were detected in multiple Doppler velocity windows for some ARs (see for example AR 11066 in Table~\ref{tab:ar_profile_type}), we then attempted to define a method which could make use of the collective behaviour of multiple Doppler velocity windows when searching for pre-emergence signatures. Our initial method is based around the identification of concurrent inflection points identified across multiple Doppler velocity windows. Such inflection points were found in 10 ARs across a wide range of longitudes (-71.5 -- 8$^{\circ}$) and latitudes (-28.3 -- 22.3$^{\circ}$). The mean time between the inflection points and $T_{FE}$, defined as $\Delta T_I$, was $100.8$ minutes, nearly double the lead time from the thresholding method. We also studied whether there was a dependence between the longitude at which an AR emerged and the Doppler velocity windows which were positive during $T_{DEI}$. In Fig.~\ref{fig:lon_vs_ar}, we plot the results of this analysis, finding ARs which emerged at less than $30^\circ$ displayed centre trends (Doppler velocity windows $<\pm$[0.3, 0.6) km s$^{-1}$) whilst ARs which emerged at higher angles displayed wing trends. 

Whilst we have been able to use inflection points for determination here, we do note several downfalls in this method that other authors should be aware of. In a live system, for example, we would have to wait for sufficient data to be collected in order to form the inflection point meaning our $\Delta T_I$ values are upper limits. Although this is a disadvantage over the original threshold method, the underlying trends within the data make it a necessary compromise. Other techniques such as cross-correlation or Bayesian online change-point detection \citep{2007arXiv0710.3742P} may be able to overcome this, but at the expense of greater computational cost. Future work would require these different techniques to be bench-marked. As this method is currently applied manually, we provide a brief overview of how we interpret different signals for completeness.

\textit{I. Strong signals.} 
The first group of ARs we discuss are those that contain strong inflection points. This includes ARs 11066 and 11114, which have been previously discussed in Sec.~\ref{sec:infle}. As previously mentioned for AR 11066 we see a strong increase in multiple velocity bins peaking at an inflection point at 18:12 UT. For AR 11114, we consider the co-temporal change across almost all sets peaking at $17$:24 UT to be a strong inflection point.

\textit{II. Diverging Signals.} 
The second group is that of divergent sets within a profile which are not quite as clear as those of group I.
Here we include ARs 11072, 11076, 11400, and 11523. 
For AR 11072 we see the profiles beginning to diverge at the time selected for $T_{DEI}$.  
Before this, we see a double-peaked burst of activity starting just after 05:24 UT, and ending around 09:24 UT. 
For AR 11076, we consider this a divergent signal due to the strong activity seen around 16:24 UT across the majority of sets, however, this is tentative.
For AR 11400, when viewing both the magnetogram profiles and image sequences of the event, we notice the slow accumulation of flux that creates divergence in the Doppler velocity windows at around 10:00 UT making the use of the threshold method less than ideal in this case. 
We also notice that just before the flux emergence there exists a small and diffuse bipole that is rapidly separating at the site from which the main flux emerges. 
As there is a clear large-scale flux emergence later, we take the closest time at which a magnetic profile exceeds the 1-$\sigma$ threshold to the larger scale emergence as $T_{FE}$. As the relationship between the activity of the smaller pre-existing flux and the larger newly-emerging flux is unclear, we tentative take $T_{DEI}$ as the inflection point which occurs just prior to the latter at 14:36 UT.
Finally, for AR 11523 we can see a great deal of activity within the Doppler profile after an initial `quiet' period. 
Whilst we must take care interpreting these profiles due to the AR $\mu$-angle (see Table~\ref{tab:ar_list}), we do not see a comparative level of activity within the magnetic profiles until around $T_{DEI}$ at 11:12 UT making it hard to determine the source of these velocities. 

\textit{III. Converging Signals.} 
The next group is that of converging signals. Here we include ARs 11088, 11130, and 11142. For AR 11088, we see a tentative convergence across a majority of sets starting around the same time of 02:48 UT, identified here as $T_{DEI}$. This convergence, rather than the apparent slower decay seen in the peaks just before flux emergence, suggests the disruption of the process that caused the activity. For AR 11130, we see a prolonged divergence converging at 05:24 UT just before emergence. Finally, for AR 11142 we once more see the convergence in the Doppler profile at around 01:48 UT before emergence. However, this convergence is significantly earlier than our other examples, and is further confounded by some minor flux emergence that occurs just before emergence.

\textit{IV. Tentative signals.} 
We consider AR 11116 a tentative signal rather than a null due to a number of minor differences between the behaviour sets around $T_{DEI}$. Firstly, we point to the tight convergence of the sets at the $T_{DEI}$, excluding the $\pm$[0.8, 1.1)~km~s$^{-1}$ set which increases from inactivity. Careful examination of the other bursts shows a lack of such convergence afterwards, in addition to the less co-temporal activity across sets. We include this result here in case future analysis can provide greater insight into this event.

\textit{V. Null signals.} We have six null results, for which no inflection point could be determined in the Doppler profiles. Null results are denoted by dashes in the $\Delta T_I$ column of Table~\ref{tab:ar_list}.

As mentioned previously we do not distinguish between the upwards velocity of the emerging flux, V$_z$, and the horizontal velocity of the divergent flow, V$_h$ to use the nomenclature of \citet{2012ApJ...751..154T} (see their Fig.~8). 
This is based on an implicit assumption that the emerging flux tube would have either recoverable components of both V$_z$ and V$_h$ at all angles or that the disruption would cause a sufficient transient change to the underlying distribution as to be measurable in the appropriate LOS component. 
A key question therefore is whether V$_z$ is itself recoverable. 
We believe the answer to be yes, due to the number of ARs with longitudes $<|20|^{\circ}$ we have been able to recover a signal  with both methods used here (e.g AR 11114 (Fig.~\ref{fig:sig_dop_profiles_smo}b)).
This is an improvement over the work of \citet{2012ApJ...751..154T,2014ApJ...794...19T}, who were not able to find signatures within $|20|^{\circ}$, though this may be due to their chosen sample. 
This is an important step towards being able to predict AR emergence with a single method across the whole disk, and not just away from the disk centre. 
Furthermore, we have been able to find a clear signature of emergence at an extreme angle (AR 11142 at 71.5$^{\circ}$), further widening the applicable window using these techniques.

In addition, we must also be aware that we may not only be recovering the pre-emergence signals of the flux that becomes the AR.
This is something that is not usually considered by other authors investigating this subject who only consider the emergence of a single flux rope into the quiet-sun,  as opposed to a flux rope emerging into another emerging flux rope or into a pre-existing active region.
Fortunately, by comparing image sequences of the magnetograms to their Doppler profiles we are able to distinguish which are the relevant signals for us to consider here.
In future work, a larger sample of ARs may allow us to automatically distinguish which signals correspond to that of the large-scale pre-emergence that we have focussed in this work.

\section{Conclusions} \label{sec:concs}

We have analysed 16 ARs to determine the time of pre-emergence signatures using Doppler velocity data from the SDO/HMI instrument. 
The studied ARs cover a large range of longitudes -71.5 $\leq \theta \leq$ 32.3.
We initially used a threshold based method to find the time of magnetic emergence, $T_{FE}$, in the [200,~300) G window (where possible).
To analyse the LOS velocities, we used two different methods, namely the thresholding method used by \citet{2014ApJ...794...19T} and the inflection point method developed here. 
In addition, this work has generalised the method of \citet{2012ApJ...751..154T} to other cadences and Doppler velocity windows. 
A comparison of the thresholding method applied to 12-minute cadence data to the results of \citet{2014ApJ...794...19T} found comparable results with $37.5$ \% of ARs displaying evidence of HDFs, with a mean $\Delta T_T$ of $58$ minutes.
For the inflection point method, we found at least tentative signals for 62.5\% of ARs, with a mean $\Delta T_I$ of 100.8 minutes.
This is higher than the mean presented by \citet{2014ApJ...794...19T}. 
When comparing the same events we find that our method finds times earlier than thresholding for the Doppler profiles.
However, future statistical work is necessary to further constrain the cases in which by-eye analysis is currently required.

\section*{Acknowledgements}
We acknowledge support from the Science and Technology Facilities Council (STFC) for the support received through grant numbers ST/P000304/1 \& ST/T00021X/1. C.J.N. acknowledges support from the European Space Agency (ESA) as an ESA Research Fellow. SDO/HMI data are provided courtesy of NASA/SDO and the HMI science team.

\bibliography{paper_refs.bib}{}
\bibliographystyle{aasjournal}

\end{document}